\documentclass[aps,prd,superscriptaddress,showkeys,showpacs,twocolumn,longbibliography,nofootinbib,floatfix]{revtex4-1}
\usepackage[utf8]{inputenc}
\usepackage{amsmath}
\usepackage{amsfonts}
\usepackage{amsthm}
\usepackage{amssymb}
\usepackage{float}
\usepackage{braket}
\usepackage{graphicx}
\usepackage{xcolor}
\usepackage{url}
\usepackage[colorlinks=true, pdfstartview=FitV, linkcolor=red, citecolor=blue, urlcolor=blue]{hyperref}
\usepackage{slashed}
\usepackage[normalem]{ulem}

\usepackage[capitalise]{cleveref}
\usepackage{placeins}

\graphicspath{{./figures/}}

\newcommand{\be}{\begin{equation}}
\newcommand{\ee}{\end{equation}}
\newcommand{\beq}{\begin{equation}}
\newcommand{\eeq}{\end{equation}}
\newcommand{\bea}{\begin{eqnarray}}
\newcommand{\eea}{\end{eqnarray}}

\newcommand{\tr}{\mathrm{tr}}

\newcommand{\Tr}{\mathrm{Tr}}

\newcommand{\Lext}[1]{L_{\mathrm{ext},#1}}
\newcommand{\jext}{j_{\mathrm{ext}}}

\renewcommand{\hat}{\widehat}

\begin{document}
\title{Thermalization from quantum entanglement: \\ jet simulations in the massive Schwinger model}

\author{Adrien Florio}
\email[]{aflorio@physik.uni-bielefeld.de}
\affiliation{Fakultät für Physik, Universität Bielefeld, D-33615 Bielefeld, Germany}\affiliation{Department of Physics, Brookhaven National Laboratory, Upton, New York 11973-5000, USA}

\author{David Frenklakh}
\email[]{dfrenklak@bnl.gov}
\affiliation{Department of Physics, Brookhaven National Laboratory, Upton, New York 11973-5000, USA}

\author{Sebastian Grieninger}
\email[]{sebastian.grieninger@stonybrook.edu}
\affiliation{Center for Nuclear Theory, Department of Physics and Astronomy, Stony Brook University, Stony Brook, New York 11794-3800, USA}

\author{\mbox{Dmitri E. Kharzeev}}
\email[]{dmitri.kharzeev@stonybrook.edu}
\affiliation{Center for Nuclear Theory, Department of Physics and Astronomy, Stony Brook University, Stony Brook, New York 11794-3800, USA}
\affiliation{Energy and Photon Sciences Directorate, Condensed Matter and Materials Sciences Division,
Brookhaven National Laboratory, Upton, New York 11973-5000, USA}

\author{Andrea Palermo}
\email[]{andrea.palermo@stonybrook.edu}
\affiliation{Center for Nuclear Theory, Department of Physics and Astronomy, Stony Brook University, Stony Brook, New York 11794-3800, USA}

\author{Shuzhe Shi}
\email[]{shuzhe-shi@tsinghua.edu.cn}
\affiliation{Department of Physics, Tsinghua University, Beijing 100084, China}
\affiliation{State Key Laboratory of Low-Dimensional Quantum Physics, Tsinghua University, Beijing 100084, China}

\bibliographystyle{unsrt}

\begin{abstract}
We investigate the emergence of thermalization in a quantum field-theoretic model mimicking the production of jets in QCD -- the massive Schwinger model coupled to external sources.  Specifically, we compute the expectation values of local operators as functions of time and compare them to their thermal counterparts, quantify the overlap between the evolving density matrix and the thermal one, and compare the dynamics of the energy-momentum tensor to predictions from relativistic hydrodynamics. Through these studies, we find that the system approaches thermalization at late times and elucidate the mechanisms by which quantum entanglement drives thermalization in closed field-theoretic systems. Our results show how thermodynamic behavior emerges in real time from unitary quantum dynamics.
\end{abstract}

\maketitle


\section{Introduction}

Understanding how isolated quantum systems evolve toward thermal equilibrium is a central question in modern theoretical physics. In quantum field theory (QFT), this process — known as thermalization — is deeply intertwined with the structure and dynamics of quantum entanglement. While thermalization traditionally refers to the emergence of thermodynamic behavior from microscopic dynamics, recent advances have highlighted the crucial role of entanglement in this process, suggesting that thermal properties can emerge locally in subsystems purely through unitary evolution, without any coupling to an external bath, see~\cite{DAlessio:2015qtq, borgonovi2016quantum} for reviews.

This entanglement-driven perspective is supported by studies of quantum quenches~\cite{Calabrese_2005, kaufman2016quantum}, holographic duality~\cite{Ryu:2006bv, Nishioka:2009un, takayanagi2012entanglement}, and the eigenstate thermalization hypothesis~\cite{PhysRevA.43.2046, Srednicki:1994mfb}, all of which point to entanglement as a key mechanism underlying the apparent loss of information and emergence of statistical behavior in closed quantum systems. In particular, the spread of entanglement entropy in time-evolving QFTs mirrors the growth of thermodynamic entropy, raising profound questions about the nature of thermalization in relativistic quantum systems and its relation to quantum information. Similar behavior have been confirmed in a variety of field theoretic models~\cite{Berges:2017hne, Berges:2017zws, Mueller:2021gxd, Desaules:2022ibp, Mueller:2024mmk, Yao:2023pht, Chen:2024pee}.

Despite these insights, a comprehensive understanding of how entanglement governs thermalization across different regimes of QFT remains incomplete. For example, the role of entanglement in strongly coupled versus weakly coupled theories, and in integrable versus chaotic systems, is still under active investigation. 

In our previous papers~\cite{Florio:2023dke, Florio:2024aix}, we addressed the real-time dynamics of thermalization using the setup that mimics the production of jets in Quantum Chromodynamics (QCD). Namely, we considered massive Schwinger model coupled to external sources that appear at some moment in time, and separate along the light cone. This setup was originally proposed by Casher, Kogut, and Susskind~\cite{Casher:1974vf}. For the (integrable) massless case, the exact solution of this model was found in~\cite{Loshaj:2011jx}. However, the most interesting case of the (chaotic) massive Schwinger model has to be studied numerically, and this can be done using the Hamiltonian real-time simulations. 

These simulations reveal a number of interesting features of the model. In particular, if we consider only a part of the produced quantum system and focus on the entanglement with its complement, we find that {\it i}) the entanglement spectrum at late times becomes very dense, suggesting an approach to a maximally entangled state in a sub-space of the full Hilbert space; and {\it ii}) the entanglement entropy at early times follows the area law (expected for the ground state in a gapped theory), but at late times approaches the volume law, as expected for a thermal state. 
These results hint at an approach to thermalization, but do not yet firmly establish it.

In this work, we significantly extend our previous results with the goal of elucidating the possible emergence of thermalization in our field-theoretic model. Specifically, we evaluate the expectation values of local operators as a function of time and compare them to the thermal expectation values, compute the overlap between the density matrix of the produced state and the thermal one, and compare the evolution of the expectation value of the energy-momentum tensor to the one predicted by relativistic hydrodynamics. These different measurements all point towards the emergence of an underlying thermodynamic description.  Our goal is to clarify the mechanisms by which entanglement leads to thermal equilibrium in field-theoretic systems and to provide a framework that connects quantum information-theoretic insights with traditional thermodynamic intuition.

This paper is organized as follows. In Sec.~\ref{sec:prelim} we review the key properties of Schwinger model and its' lattice formulation and explain how we perform numerical simulations. In Sec.~\ref{sec:local_obs} we study the effective temperature emerging from different local observables. In Sec.~\ref{sec:entropy} we discuss finding the effective temperature from the entanglement entropy of an interval in the middle of the system and from thermodynamic entropy. In Sec.~\ref{sec:overlap} we compute the overlap between the reduced density matrix in the jet system and a thermal density matrix to extract an effective temperature from the entire state at once. In Sec.~\ref{sec:hydro} we study the system behavior after removing the external sources using the equations of ideal relativistic hydrodynamics. Finally, in Sec.~\ref{sec:conclusion} we provide a conclusion and outlook. Seven appendices provide technical details and add more perspective on some of the questions discussed in the main text.

\section{Lattice formulation of the Schwinger model} \label{sec:prelim}

Quantum electrodynamics in 1+1 spacetime dimensions, known as the Schwinger model, is often used as a toy model for studying phenomena of interest in QCD, including jet fragmentation \cite{Casher:1974vf, Kharzeev:2012re, Loshaj:2011jx, Florio:2023dke, Florio:2024aix}. The idea, originally due to Casher, Kogut and Susskind \cite{Casher:1974vf}, is to model the jets with two high-energy external sources that propagate back to back and couple to the theory via the electric field they produce. This electric field is dynamically screened by quark-antiquark pair creation, that resembles fragmentation of jets. In the Schwinger model this real-time process can be explored either analytically (in the  massless fermion limit) \cite{Loshaj:2011jx,Kharzeev:2012re} or numerically with Hamiltonian simulation methods \cite{Florio:2023dke, Florio:2024aix, Janik:2025bbz}. In the Hamiltonian formulation the continuum version of the theory is given by
\begin{align}
H &=\int dx \left [ \frac{1}{2}E^2+\bar{\psi}(-i\gamma^1\partial_1+g\gamma^1A_1+m)\psi +\jext^1 A_1 \right ], \label{eq:cont_Ham}
\end{align}
where $\psi$ is the single-flavor two-component fermion field, $A_1$ is the $U(1)$ gauge potential\footnote{We have fixed the temporal gauge $A^0 = 0$.} and $E = F_{01}$ is the electric field strength. $m$ and $g$ are the fermion mass and electric charge, respectively; note that the latter has mass dimension 1 in 1+1 dimensions. $\jext^1$ is the external source modeling two jets propagating back to back along the lightcone, chosen as
\begin{equation}
    \jext^1(x,t) = g[\delta(x-t) + \delta(x+t)]\theta(t) \, , \label{eq:jext}
\end{equation}
where we have set the point where the jets first appear as the origin of spacetime.

To simulate the system numerically we have to discretize it; to this end we use staggered fermions $\chi_i$ \cite{Banks:1975gq} and completely integrate out the gauge field using Gauss's law which is possible for open boundary conditions\footnote{We work without a constant background electric field, in other words $\theta=0$.}. We arrive at the following Hamiltonian in the fermionic basis:
\begin{align}
    H(t) &= H_{kin} + H_m + H_E(t) \,, \label{eq:lat_Ham} \\
    H_{kin} &= -\frac{i}{2a}\sum_{n=1}^{N-1}
\big(\chi^\dag_{n}\chi_{n+1}-\chi^\dag_{n+1}\chi_{n}\big) \, , \\
    H_m &= m\sum_{n=1}^{N} (-1)^n \chi^\dag_n\chi_n  \, , \\
    H_E(t) &= \frac{ag^2}{2}\sum_{n=1}^{N-1} (L_n + \Lext{n}(t))^2\, . 
\end{align}
Here, $a$ is the lattice spacing and $m$ and $g$ are lattice mass and coupling parameters which are not the same as the continuum parameters in Eq. (\ref{eq:cont_Ham}). $L_n$ is the dynamical electric field on the $n$-th link which may be expressed with the help of Gauss's law as:
\begin{equation} \label{eq:Gauss_law}
    L_n -L_{n-1} = Q_n \Rightarrow L_{n} = \sum_{i = 1}^n Q_i \,,
\end{equation}
using the charge operator $Q_i = \chi^\dagger_i \chi_i + \frac{(-1)^i-1}{2}$. The time-dependent electric field, produced by the external sources (\ref{eq:jext}), reads
\begin{equation}
    \Lext{n}(t) = -\theta\left(\frac{t}{a} - \left|n-\frac{N}{2}\right|\right),
\end{equation}
and it plays the role of a quench in our system. Furthermore, the Hamiltonian can be mapped to a spin basis via the Jordan-Wigner transformation, see Appendix~\ref{app:spin} for a short discussion and Refs.~\cite{Florio:2023dke, Florio:2024aix} for more details.

In this work, we will encounter the following observables 
\begin{equation}
    \hat {\cal O} = \{\widehat{\bar\psi\psi}_n, \widehat{\bar\psi\psi}_n\widehat{\bar\psi\psi}_{n+1}, \widehat K_n ~ \mathrm{where}~ n\in[\frac{N-l}{2},\frac{N+l-2}{2}]\}\, , \label{eq:O_set}
\end{equation}
with  the chiral condensate and kinetic energy density operators given by, respectively
\begin{align}
    \widehat{\bar\psi\psi}_n &= \frac{(-1)^n}{a}\big(\chi_n^\dag\chi_n-\frac{1}{2}\big)\, , \label{eq:cond_def} \\
    \widehat K_{n} &= -\frac{i}{2a}
\big(\chi^\dag_{n}\chi_{n+1}-\chi^\dag_{n+1}\chi_{n}\big)\, , \label{eq:kin_def}
\end{align}
and where the hat indicates that we are dealing with quantum operators. $l$ denotes the size of a representative sample of sites sufficiently far from the boundaries.

To simulate jet fragmentation, we begin by initializing the system in the ground state of the unquenched Hamiltonian $H(t=0)$. Afterwards, we perform unitary time evolution with the time-dependent Hamiltonian~\eqref{eq:lat_Ham}. We are dealing with systems where the size of Hilbert space is too large for a full diagonalization of the Hamiltonian, thus we choose tensor network methods to simulate jet fragmentation. Namely, we use a matrix product state (MPS) representation of the quantum state with density matrix renormalization group (DMRG) method for ground state preparation \cite{White:1993zza, Schollwock:2005zz} and time-dependent variational principle (TDVP) for time evolution \cite{Haegeman:2011zz, Haegeman:2016gfj}. These methods have been successfully applied to the Schwinger model previously to study the spectrum, thermal properties and real-time dynamics, see e.g. \cite{Banuls:2013jaa, Buyens:2013yza, Banuls:2015sta, Papaefstathiou:2024zsu}.  We use the iTensor Julia library~\cite{itensor, itensor-r0.3} for all tensor network calculations. Due to growth of entanglement during jet fragmentation the bond dimension of the MPS increases with time, setting an upper bound $t_{max}$ on how far one can simulate the evolution. This implies a natural bound on the system size, as the jets propagate at the speed of light, so it is natural to use a system of size $N\approx\dfrac{2 \,t_{max}}{a}$. We find that the value of $t_{max}$ depends on the fermion mass $m$: for smaller fermion mass, pair production is stronger and therefore entanglement growth is faster, thus reducing the time interval accessible to a reliable tensor network simulation. 

To compute reference thermal expectation values, we use a combination of Exact-Diagonalization computations and our ITensor-based implementation of the purification algorithm~\cite{Verstraete:2004gdw, Zwolak:2004nwu, Feiguin:2005jud}. Detailed explanations and numerical extrapolations are presented in Appendix~\ref{sec:thermal_prelim}. Throughout the paper we express the temperature and other dimensionful quantities in terms of physical units, defined by the mass of the first excited state in the Schwinger model, namely the pseudoscalar meson. We describe how to find its mass, denoted by $M_s$, and perform an infinite volume extrapolation for it in Appendix~\ref{app:meson}.

In previous work \cite{Florio:2024aix}, we observed equilibration of local observables (chiral condensate and electric field energy) in the central region of the system at late times. In the subsequent sections, we will present strong evidence that the system reaches a thermal state. 

\section{Thermalization from the local observables} \label{sec:local_obs}

\begin{figure}
    \centering
    \includegraphics[width=\linewidth]{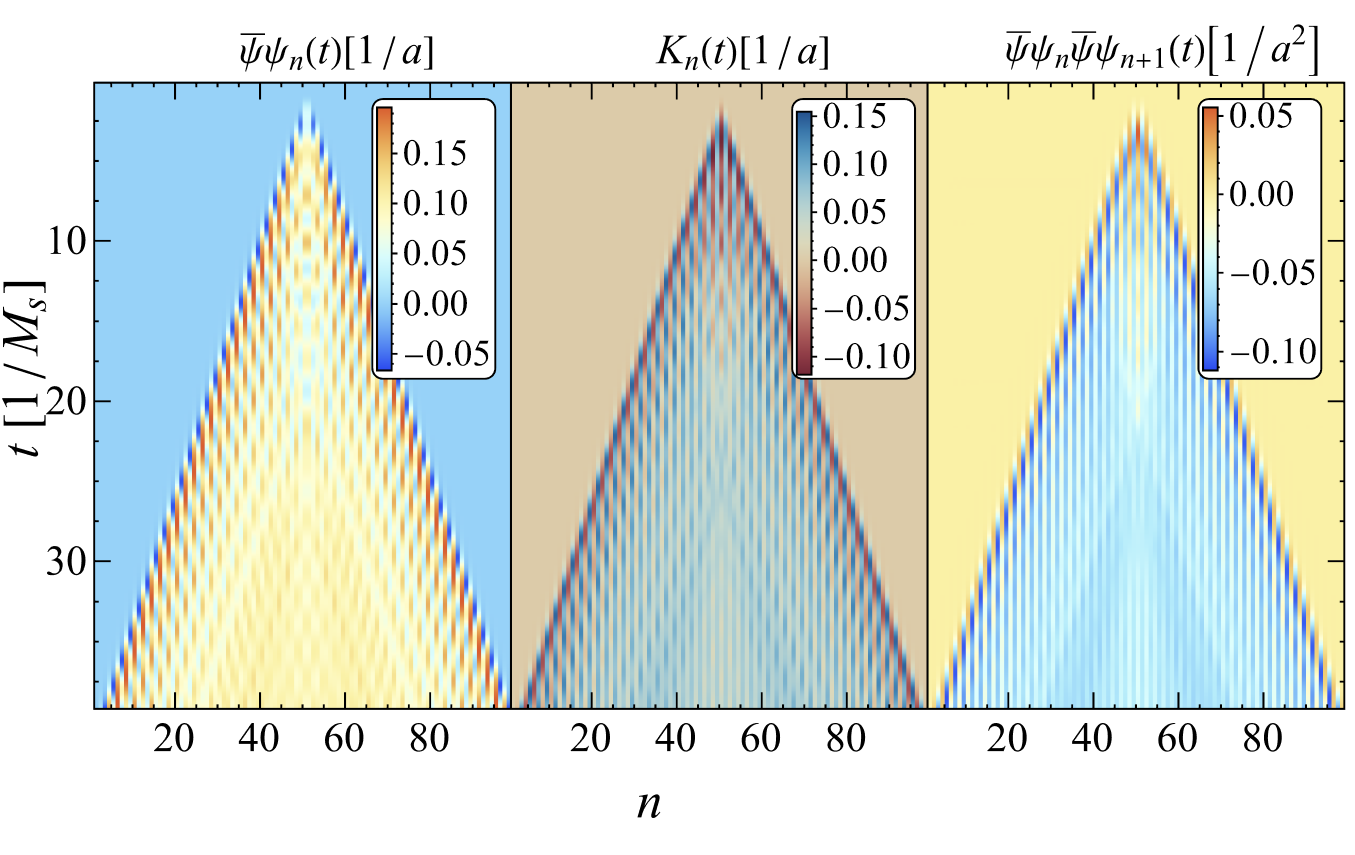}
    \caption{Spacetime dependence of the local chiral condensate (left), kinetic energy density (center) and nearest-neighbor chiral condensate two-point function (right) in a simulation with $N=100, g=0.5/a, m=0.5\,g$. Equilibration in the central region is evident, up to fluctuations between even and odd sites due to the staggering effects. Vacuum expectation values are subtracted.}
    \label{fig:obs_density}
\end{figure}

As explained in Section~\ref{sec:prelim}, from the real-time simulation we obtain the time-dependent wavefunction
\begin{equation}
    |\Psi(t)\rangle = {\cal T} e^{-i \int_0^t dt'H(t')} |\Psi(0)\rangle \, ,
\end{equation}
that can be used to find the expectation values of local observables. For the purpose of extracting an effective temperature, we compute the time-dependent expectation values of operators belonging to the set (\ref{eq:O_set}), namely chiral condensate, condensate 2-point function and kinetic energy density, as:
\begin{align}
    {\cal O}(t) &= \langle \Psi(t)| \hat{\cal O}|\Psi(t)\rangle \, , \label{eq:O(t)_def} 
\end{align}
We find that each of these local observables reaches an equilibrated distribution in the central part of the system, as shown in Fig. \ref{fig:obs_density}.

\begin{figure}
    \centering
    \includegraphics[width=1\linewidth]{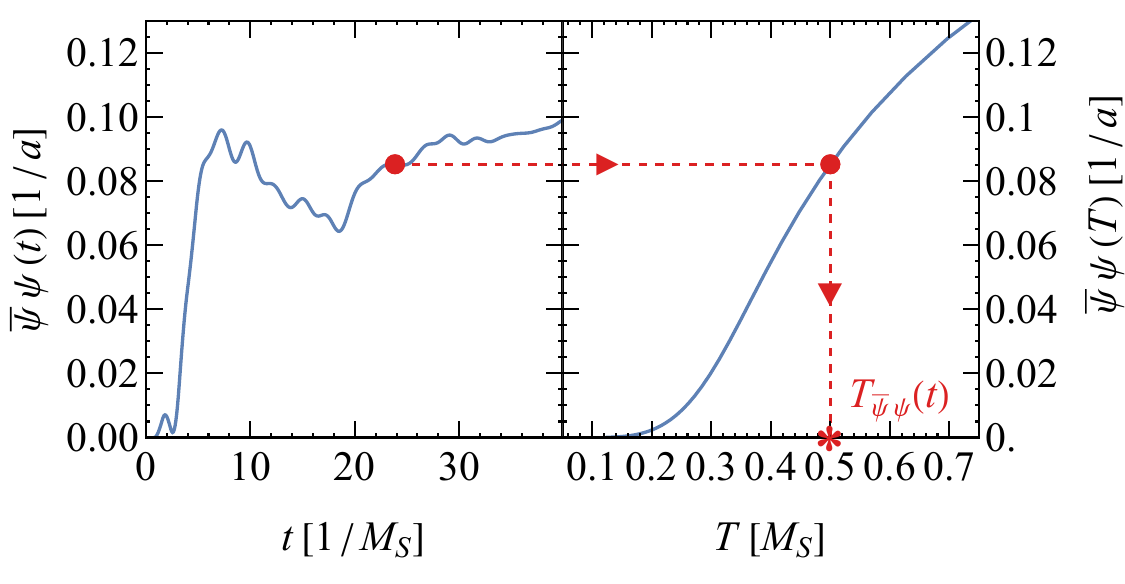}
    \caption{Visualization of the procedure to find the effective temperature as a function of time from the time dependence of a local observable, in the present case the chiral condensate averaged over 10 lattice sites at the center of the system.}
    \label{fig:T_extraction}
\end{figure}

Comparing these local expectation values to the temperature dependence of each of the observables $\mathcal{O}$ (see Appendix~\ref{sec:thermal_prelim} for details on extracting the thermal expectation values, and in particular Fig.~\ref{fig:local_obs_vs_T} therein) allows us to infer an effective temperature $T_{\cal O}(t)$. This procedure is illustrated in Fig. \ref{fig:T_extraction} in the case of chiral condensate used as the observable. Notice that this prescription always assigns an effective temperature to the time-dependent expectation value of each operator, even if the state is far from thermal. To achieve thermalization and interpret the effective temperature as a physical temperature of the subsystem, we require that all effective temperatures extracted from different observables be the same. 

To mitigate the effect of random fluctuations, we average the local observables over $L/2$ physical lattice sites at the center of the system. We experimented with different values of $L$; all the results reported in this section correspond to a representative value $L=10$. The site-to-site fluctuations allow us to estimate the uncertainty of the observable. For any of the observables from the set (\ref{eq:O_set}) at a physical site (summed over the two corresponding staggered sites) denoted as ${\cal O}_n^{\rm phys}$, we define 
\begin{align}
    \bar{\cal O} &\equiv \frac{2}{L}\sum_{n = 1}^L {\cal O}_n^{\rm phys} \, , \\
    \Delta{\cal O} &\equiv \sqrt{\frac{2}{L(L/2-1)}\sum_{n=1}^L({\cal O}_n^{\rm phys} - \bar{\cal O})^2} \, ,
\end{align}
where the second equation defines the standard error for a sample. We evaluate the mean temperature as well as the upper and lower limits of the uncertainty interval as $T_{\cal O}(\bar{\cal O})$ , $T_{\cal O}(\bar{\cal O} + \Delta{\cal O})$, $T_{\cal O}(\bar{\cal O} - \Delta{\cal O})$, correspondingly.

\begin{figure}
    \centering
    \includegraphics[width=1\linewidth]{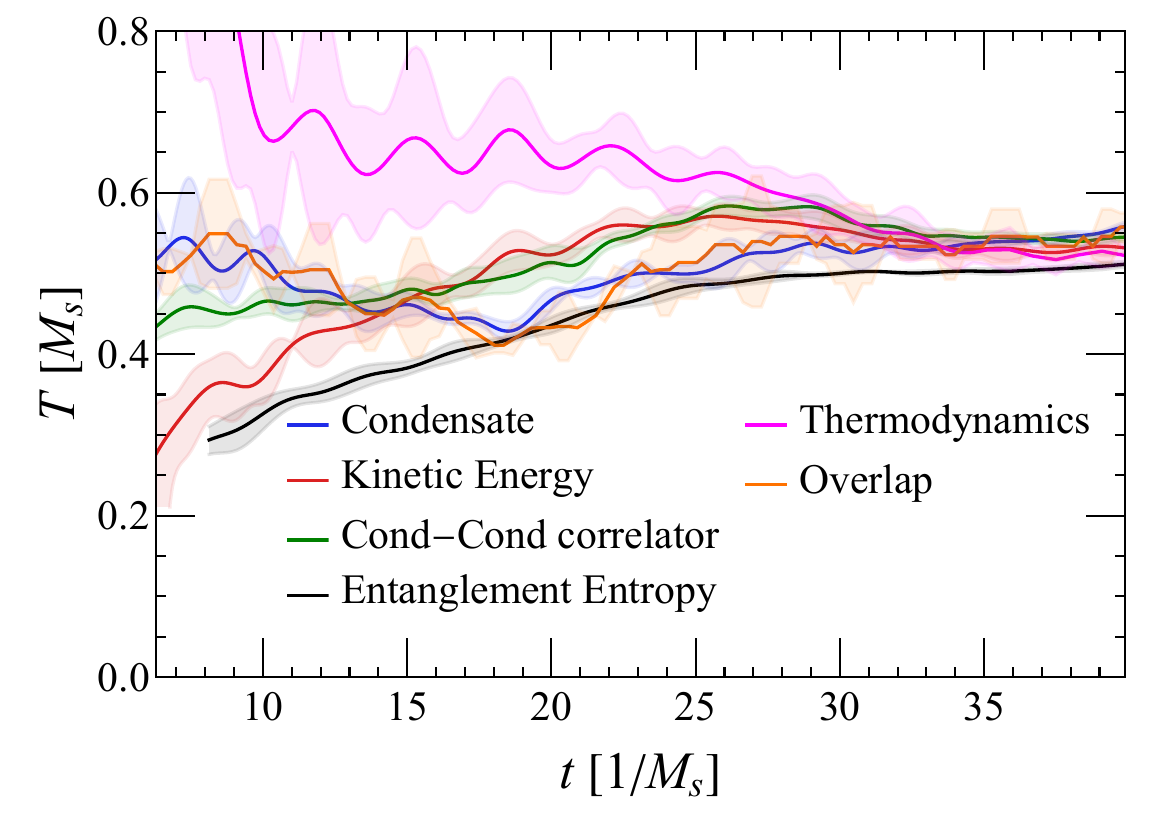}
    \caption{Effective temperatures in the jet simulation found from the local observables, entanglement entropy, thermodynamic relation~\eqref{eq:thermo_temp} and maximizing the overlap between the jet reduced density matrix and a thermal density matrix. The procedure to evaluate temperature from the chiral condensate, kinetic energy and condensate-condensate correlator is described in Sec.~\ref{sec:local_obs}. Extraction of the effective temperature from the entanglement entropy and the thermodynamic relation (\ref{eq:thermo_temp}) are described in Sec. \ref{sec:entropy}. Finding the temperature by overlapping density matrices is discussed in Section \ref{sec:overlap}. Fermion coupling and mass are $g=0.5/a$,  $m=0.5\,g$. Units of temperature and time are derived from the meson mass $M_S$ corresponding to the energy of the first excited state in the infinite volume limit (see Appendix~\ref{app:meson} for details).}
    \label{fig:thermalization_m025}
\end{figure}

The time dependence of the effective temperature extracted from the three local observables for the system with parameters $g = 0.5/a, m = 0.5\,g$ is displayed in Fig.~\ref{fig:thermalization_m025} along with temperature estimates based on the entanglement entropy, which are described in Sec.~\ref{sec:entropy}, and the temperature of the state with the greatest overlap with the jet reduced density matrix, discussed in Sec.~\ref{sec:overlap}. After time $t\approx30/M_s$, the temperature extracted from various local observables  is in agreement between observables within the error bars, around $T/M_s= 0.53\pm 0.02$. Furthermore, as will be discussed in Sec.~\ref{sec:entropy} and \ref{sec:overlap}, the measures that go beyond local observables converge to the same temperature values. 

\begin{figure*}
    \centering
    \includegraphics[width=0.75\textwidth]{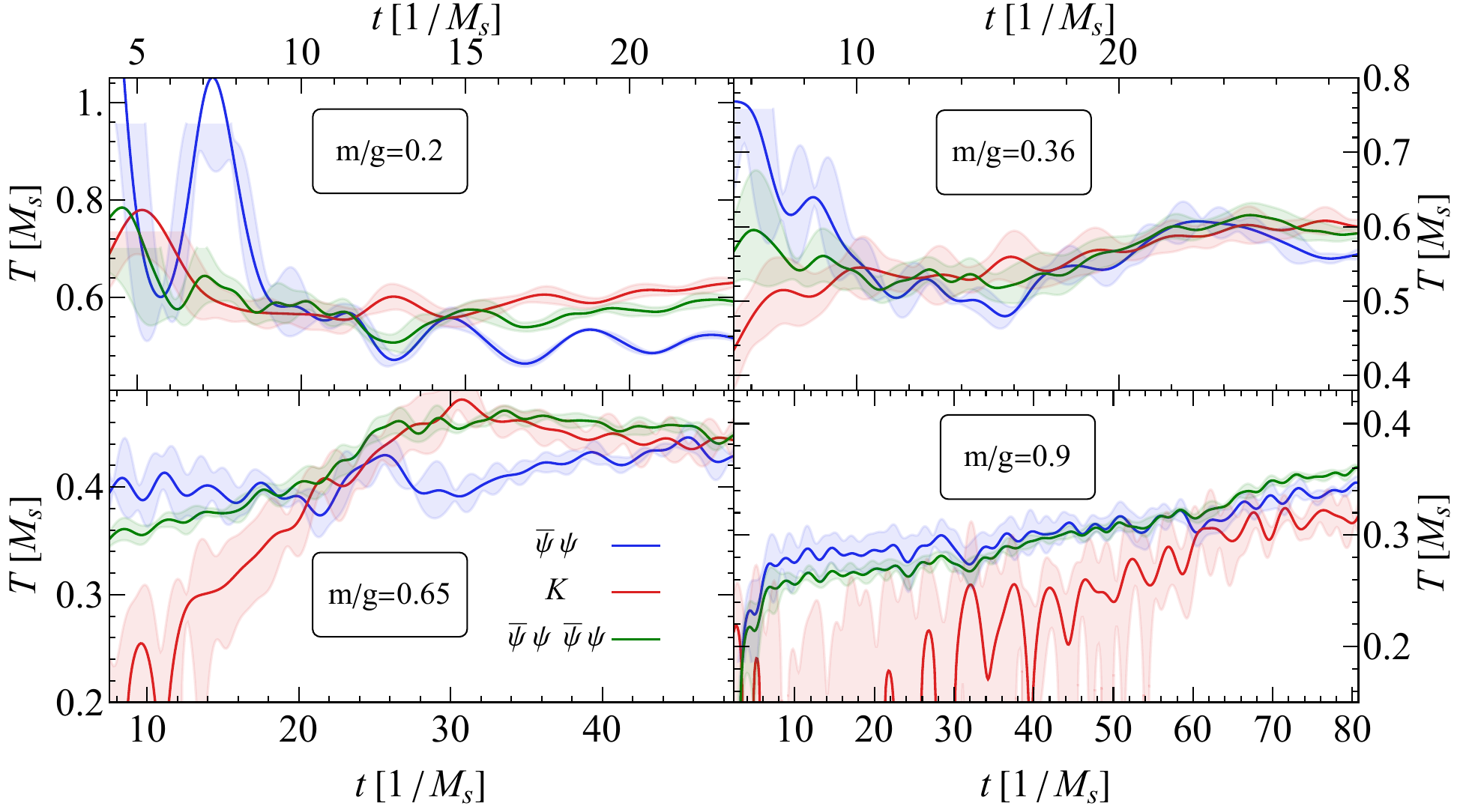}
    \caption{Comparison of the dynamics of thermalization in the jet simulation from local observables at several values of fermion mass $m$. Temperature and time are expressed in the physical units, namely the pseudoscalar meson mass, as explained in the text.  Fermion coupling is $g=0.5/a$ in all cases. Thermalization at large fermion mass, $m/g=0.9$, is slower compared to $m/g\sim 0.5$. For small fermion mass, $m/g=0.2$, due to computational cost we could not go to comparable physical times, however in the available time interval the observables equilibrate to slightly different effective temperatures, indicating a lesser degree of thermalization compared to $m/g\sim 0.5$.}
    \label{fig:thermalization_mass_scan}
\end{figure*}

The parameters of the system, namely the fermion mass and coupling, can have an effect on the effective temperature and the dynamics of thermalization, for instance the thermalization time. We study thermalization at the level of local observables for a variety of fermion mass parameters\footnote{In principle, one could also vary the coupling $g$, but this would be equivalent to varying the lattice spacing $a$. Although it is meaningful to vary $a$ and study discretization effects by going towards the continuum limit we leave it to future work.} , see Fig.~\ref{fig:thermalization_mass_scan}. The time interval accessible to the simulation depends on the fermion mass. Namely, a faster pair production rate for smaller $m$ implies a faster growth of entanglement and, correspondingly, the bond dimension of an MPS representation with fixed accuracy. As one can see in Fig.~\ref{fig:thermalization_mass_scan}, none of the studied mass parameters achieves the level of thermalization we observed for $m=0.5\,g$. The values of the effective temperature extracted from various local observables are slightly different, beyond the error bars. Nevertheless, as the fermion mass increases, the effective temperature decreases across the whole set of observables. This behavior is expected as it reflects the fact that pair production is suppressed for heavy fermions.

Slower thermalization rate at very large and small values of the fermion mass is consistent with the fact that the theory becomes integrable in both limits. At very large mass, Schwinger model reduces to a free fermionic theory, while at zero mass it is related by the bosonization duality to a free massive boson theory \cite{Coleman:1975pw}. In either of these cases we do not expect thermalization to occur. Our findings reveal that $m/g = 0.5$ is at the interface of the two regimes, sufficiently far from either of the integrable regimes of the theory. However, increasing or decreasing $m/g$ from this value brings the system close to an integrable regime and we do not find a strong evidence of thermalization.

\section{Thermalization in entanglement entropy} \label{sec:entropy}

\begin{figure}
    \centering
    \includegraphics[width=\linewidth]{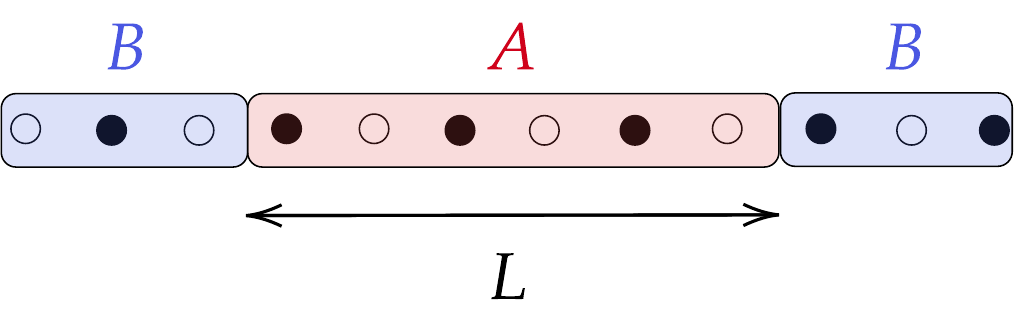}
    \caption{The bipartition of the system used to study the entanglement of the central region of length $L$ (subsystem $A$) with the complement (subsystem $B$). Subsystem $A$ is centered around the center of the system.}
    \label{fig:AB_bipartition}
\end{figure}

Rapid thermalization at early stages of a heavy-ion collision leading to emergence of quark-gluon plasma has not found an explanation in QCD yet. It has been conjectured that the mechanism of this thermalization process may be related to quantum entanglement. Our setup offers a unique opportunity to access the dynamics of entanglement entropy and thermalization at the same time, in a model closely related to QCD. 

We begin by studying the dynamics of the entanglement entropy of a subsystem consisting of  an interval at the center of the system, with its complement; this bipartition is displayed in Fig.~\ref{fig:AB_bipartition}. The entanglement entropy is defined as
\begin{align}
    S(t) = -\tr_A(\rho_A(t)\log \rho_A(t))  \, , 
\end{align}
where $\rho_A(t) = \tr_B\rho(t)$, $\rho(t) \equiv |\psi(t)\rangle\langle\psi(t)|$ and $\tr_i$ denotes tracing over the degrees of freedom of subsystem $i = A,B$. We find the spectrum of the reduced density matrix $\rho_A$, $\{\lambda_i\}$, from an MPS representation of the state in a way described in Appendix~\ref{app:entanglement_spectrum}.

\begin{figure}
    \centering
    \includegraphics[width=1\linewidth]{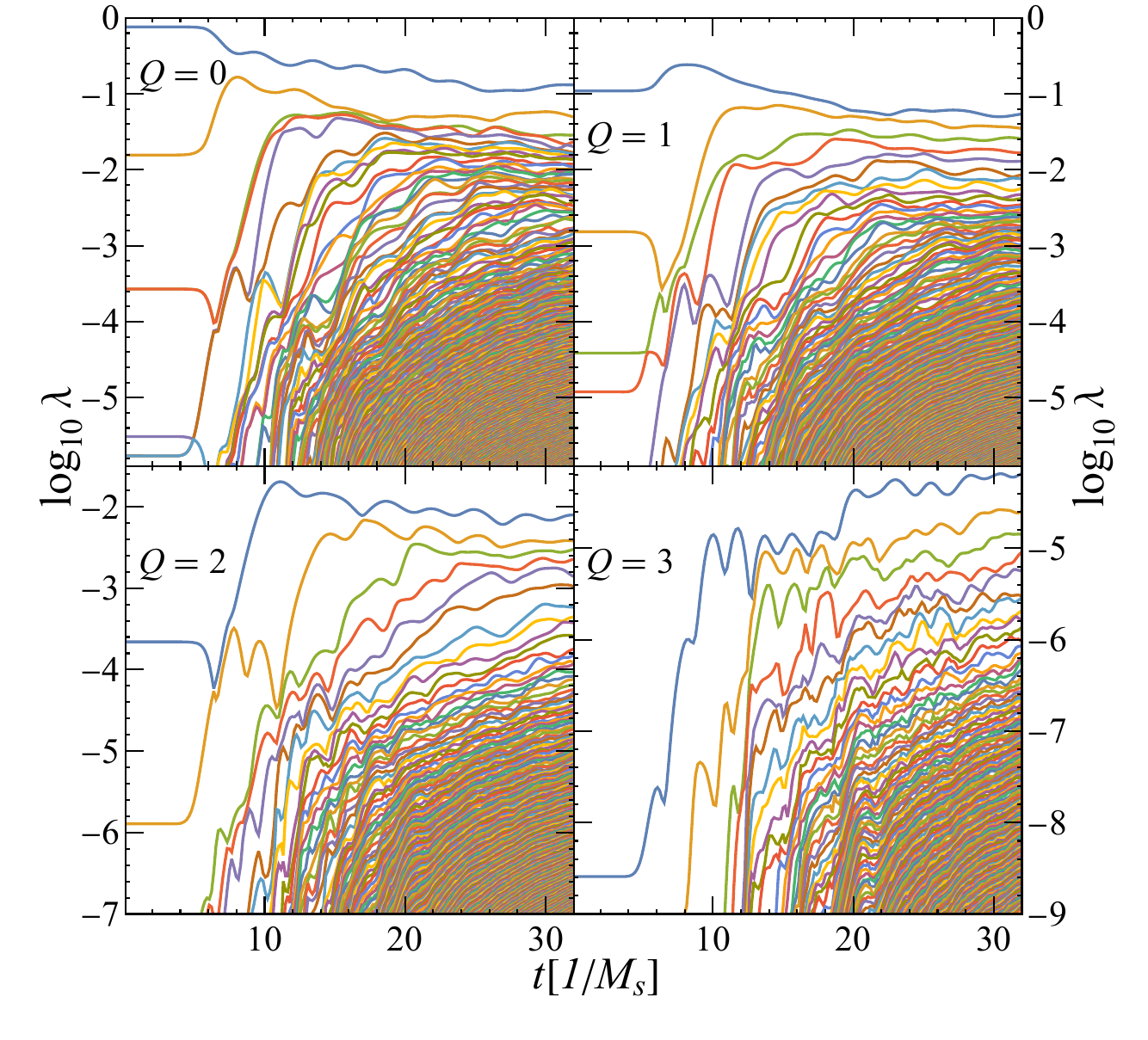}
    \caption{Entanglement spectrum in four different charge sectors for a subsystem of size $L=12$ as a function of time, shown as the logarithm for legibility. Total size of the system is $N=100$, fermion coupling and mass are $g = 0.5/a, m = 0.5 g$.}
    \label{fig:ent_spectrum}
\end{figure}

We display the entanglement spectrum in a subsystem of size $L=12$ in Fig.~\ref{fig:ent_spectrum}. Throughout this section, we consider the system with parameters $g=0.5/a, m=0.5g$ and $N=100$. Because of the presence of a conserved charge, namely the electric charge, each eigenstate of the reduced density matrix has a particular charge. In Fig.~\ref{fig:ent_spectrum} we organize the eigenvalues into 4 different panels by charge sector. In each charge sector we find that at early times, very few Schmidt states contribute to the reduced density matrix, reflecting that the subsystem is close to a pure state. At late times, many more Schmidt states emerge and the subsystem moves towards a maximally entangled state.

\begin{figure}
    \centering
    \includegraphics[width=1\linewidth]{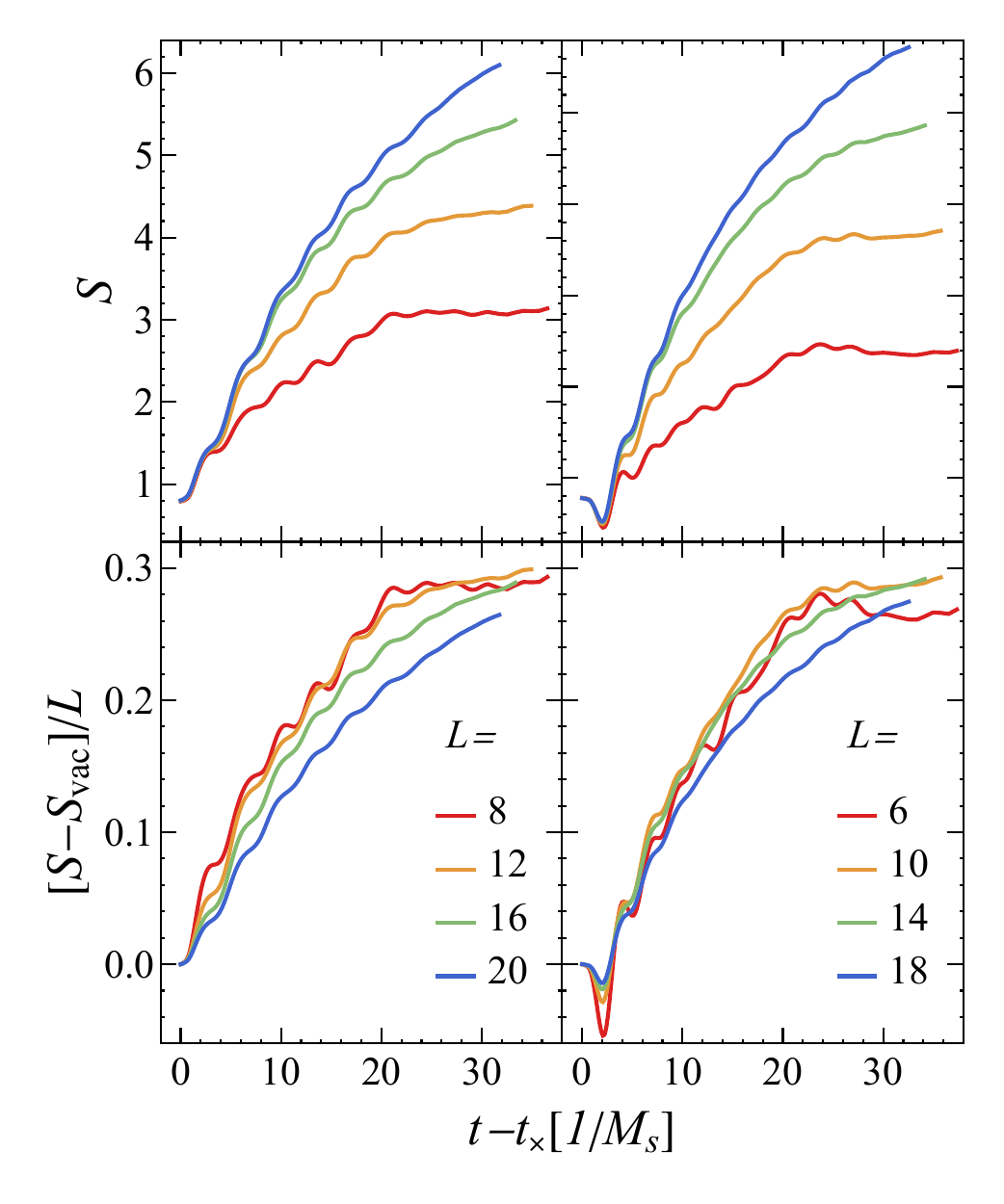}
    \caption{Top panel: time-dependence of $S(L)$ with time shifted such that the jets cross the subsystem boundary at $t-t_\times=0$. Matching of the initial shape of the curves is the signature of an area law. Values of $L$ are separated on the right and on the left panels due to staggering effects. Bottom panel: vacuum subtracted entanglement entropy per unit length, with the time shifted same as above. Different subsystem size curves reach a plateau at late times at the same value of entropy density, signifying the volume law of entanglement. System parameters are $m=0.5 g, g=0.5/a$. Subsystems with $L=2,4$ are found to be too small to obey area law and are excluded. }
    \label{fig:S_area_vol_law}
\end{figure}

From the entanglement spectrum, it is straightforward to compute entanglement entropy as 
\begin{equation}
    S(t) = -\sum_i \lambda_i \log\lambda_i. 
\end{equation}
The external charges cross the boundary between subsystems $A$ and $B$ at $t_\times = L/2$. This is the moment when the entanglement entropy starts changing. Offsetting by the jet arrival times, we find that the initial time dependence of entanglement entropy follows a universal behavior, see the top panel of Fig.~\ref{fig:S_area_vol_law}. It is a signature of the area law of entanglement, characteristic for ground state and weakly entangled states of gapped theories~\cite{Eisert:2008ur}. We find that this behavior is different for even and odd $L/2$ due to the staggering effects. After an intermediate stage of growth, the entanglement entropy saturates at late times. By subtracting the vacuum value of the entropy and rescaling by the subsystem size, we find that the entanglement entropy follows a volume law at late times, as shown in the bottom panel of Fig.~\ref{fig:S_area_vol_law}. As volume law is expected for a thermal state~\cite{Bianchi:2021aui, Nakagawa:2017yiw}, this is further evidence of thermalization.

An intriguing connection between entanglement entropy and Gibbs statistical entropy has been suggested in a number of contexts, including high-energy particle collisions. Namely, the statistical entropy of the final state has been proposed to originate from the entanglement entropy of the initial state. \cite{Baker:2017wtt, Berges:2017zws, Tu:2019ouv}. The two entropies have been found to be exactly equal  in an analytical model of pair production \cite{Florio:2021xvj, Grieninger:2023ehb, Grieninger:2023pyb}. Motivated by this connection, we attempt to interpret the previously described entanglement entropy as Gibbs entropy, and extract the effective temperature from it using the function $T_S$, inverse to $S(T)$ defined in Eq.~\eqref{eq:S(T)}. 

The uncertainty of the entropy-based effective temperature is estimated by taking a sample of entanglement entropy densities with $L\in[8,12]$\footnote{This range is chosen based on the time dependence of the entropy density. We find from Fig.~\ref{fig:S_area_vol_law} that for these values of $L$ the volume law is established towards late times and finite volume effects are not significant.} for which we  calculate the mean and standard error:
\begin{align}
    \bar s_{EE} &= \frac{1}{3}\sum_{l = 4}^6 \frac{S_{EE}(L=2l) - S_{EE}^{vac}(L=2l)}{2l} \, ,\\
    \Delta s_{EE} &= \sqrt{\frac{1}{6} \sum_{l = 4}^6 \left(\frac{S_{EE}(L=2l) - S_{EE}^{vac}(L=2l)}{2l} - \bar s_{EE}\right)^2} \, 
\end{align}

The effective temperature resulting from applying the function $T_S$ to $\bar s_{EE}$ and the corresponding uncertainty interval are shown in Fig.~\ref{fig:thermalization_m025}. It reaches a close agreement with the universal temperature extracted from the local observables, as explained in Sec.~\ref{sec:local_obs}. This is strong evidence of thermalization as well as a direct connection between the entanglement entropy and Gibbs entropy.

We can further explore this connection by recalling the basic thermodynamic relation connecting energy density $\epsilon$, pressure $P$, temperature $T$ and Gibbs entropy density $s$ in equilibrium:
\begin{equation}
    s = \frac{\epsilon + P}{T}\, . \label{eq:thermo_relation}
\end{equation}
In the jet simulation we have access to the energy density and pressure, defined by the components of the stress-energy tensor. Assuming an ideal fluid form, this reads:
\begin{equation}\label{eq:ideal tmunu}
    T^{\mu\nu} = \epsilon u^{\mu} u^\nu -P(g^{\mu\nu}-u^\mu u^\nu).
\end{equation}
In the center of the lattice, where the two-velocity is $u^\mu=(1,0)$ by symmetry, the energy density and pressure are easily identified as ~\cite{Janik:2025bbz}:
\begin{align}\label{eq:energy and pressure in the center}
    \epsilon_{N/2} &= T^{00}_{N/2} = \frac{1}{a}(K_{N/2} + m \bar\psi\psi_{N/2}) + \frac{g^2}{2}L_{N/2}^2 \, , \\
    P_{N/2} &= T^{11}_{N/2} = \frac{1}{a}K_{N/2} - \frac{g^2}{2}L_{N/2}^2 \, ,
\end{align}
where $K_n$ and $\bar\psi\psi_n$ are the local kinetic energy and chiral condensate defined in Eqs.~\eqref{eq:kin_def},~\eqref{eq:cond_def}, respectively\footnote{Notice that the computation of the energy density implies the average over a few sites. However, relativistic corrections to Eq.~\eqref{eq:energy and pressure in the center} due to the two-velocity $u^\mu$ remain small in the central region, and we neglect them.}.

As we just showed assuming a one to one correspondence between the Gibbs and entanglement entropy is consistent with the late-time thermalization in the center of the jet system. Thus, we use the entanglement entropy density in place of the Gibbs entropy density to find the effective temperature based on the thermodynamic relation (\ref{eq:thermo_relation}):
\begin{equation}
    T_{therm} =\frac{\epsilon_{N/2}+P_{N/2}}{\bar{s}_{EE}} =\frac{2 K_{N/2} + m\bar\psi\psi_{N/2}}{a \, \bar s_{EE}} \, . \label{eq:thermo_temp}
\end{equation}
The resulting temperature is presented in Fig. \ref{fig:thermalization_m025} with the uncertainties propagated from $\Delta K$, $\Delta \nu$ and $\Delta s_{EE}$. When thermalization at the level of local observables is achieved, the thermodynamic temperature reaches the same values, confirming the emergence of a thermal state and the identification between the entanglement entropy and Gibbs entropy. This is the main result of our work.

Before moving on to other observables,  let us mention that inspired by the approach to maximal entanglement in the symmetry-resolved entanglement spectrum, shown in Fig.~\ref{fig:ent_spectrum}, one can study how thermalization interplays with charge conservation. We report an early progress in this direction in Appendix~\ref{app:symmetry_resolved} but leave a detailed study of this question to future work. 

\section{Overlaps with thermal states} \label{sec:overlap}

In the previous sections, we have observed thermalization in the jet system at the level of local observables and entanglement entropy. It is then natural to ask how close the quantum state of the jet system is to a thermal state. In order to make a meaningful comparison, we will focus on a subsystem $A$ of size $L$ of the jet system (see Fig. \ref{fig:AB_bipartition}) described in terms of a reduced density matrix $\rho_A$. We measure the distance of this reduced density matrix to thermal states, characterized by density matrices $\rho_\beta$, and determine a temperature by selecting the thermal state closest to $\rho_A$. To measure the proximity of $\rho_A$ and $\rho_\beta$, different metrics can be used \cite{PhysRevA.102.012405}. Two prominent measures of the distance between states are the fidelity $F$ and the trace distance $D$ \cite{Jozsa:1994qja,Nielsen:2012yss}:
\begin{align}
    F(\rho_A, \rho_\beta) = \left(\Tr \sqrt{\sqrt{\rho_A} \,\rho_\beta \sqrt{\rho_A}}\right)^2 \,,\\
    D(\rho_A, \rho_\beta) = \Tr\left(\sqrt{(\rho_A-\rho_\beta)^2}\right).
\end{align}
However, both quantities require the calculation of square roots, which can be numerically costly and challenging with tensor network methods for larger subsystems. Hence, we use two proxies for these metrics. Instead of the fidelity, we will compute the \textit{overlap} between two mixed states as:
\begin{equation}
    f(\rho_A,\rho_\beta) \equiv \frac{\Tr (\rho_A \,\rho_\beta)}{\sqrt{\Tr\,\rho_A^2  \, \Tr\, \rho_\beta^2}}\,. \label{eq:overlap_definition}
\end{equation}
The overlap is one if $\rho_A=\rho_\beta$ and can be easily computed with tensor network methods if one has access to the reduced density matrix and the thermal density matrix in MPO form. To find an MPO of the thermal state we begin with an MPS representation of the purified thermal state. We refer the reader to Appendix \ref{app:entanglement_spectrum} for further details of our algorithm.  

As a second measure, instead of using the trace distance we consider the (square root of the) Hilbert-Schmidt (HS) distance~\cite{Vedral:1997hd}
\begin{equation}
    D_\text{HS}(\rho_A,\rho_\beta)=\sqrt{\Tr[(\rho_A-\rho_\beta)^2]}.
\end{equation}
 The advantage of this measure is that it can be computed without diagonalizing the matrices, and therefore it is often used as a cost function in learning and variational problems \cite{Travnicek:2019qvo,larose2019variational,arrasmith2019variational,Cerezo:2023nqf,Braccia_2021}. Furthermore, the HS distance bounds the trace distance from above \cite{PhysRevA.100.022103}. In what follows, instead of using the HS distance, we use its complement
\begin{equation}
    d_\text{HS}(\rho_A,\rho_\beta)=1-D_\text{HS}(\rho_A,\rho_\beta),
\end{equation}
such that $d_\text{HS}$ approaches 1 when $\rho_A\simeq \rho_\beta$.

All results presented in this section are obtained for a system with $g=0.5/a$, $m=0.5g$ and $N=100$. Our general procedure is the following: we take the jet wave function at a given time and construct the reduced density matrix of a subinterval $L$ centered around the middle. In this work, the thermal states are produced for $N=12$. For these states, we construct the thermal density matrix (of length $N=12$) which we subsequently trace down to match the subsystem size of the reduced density matrix. Then we proceed to compute the overlap and HS distance of the two objects. For a given reduced density matrix, we scan a range of temperatures and select the temperature where the normalized overlap is maximal (or the HS distance minimized).

\begin{figure}
    \centering
    \includegraphics[width=1\linewidth]{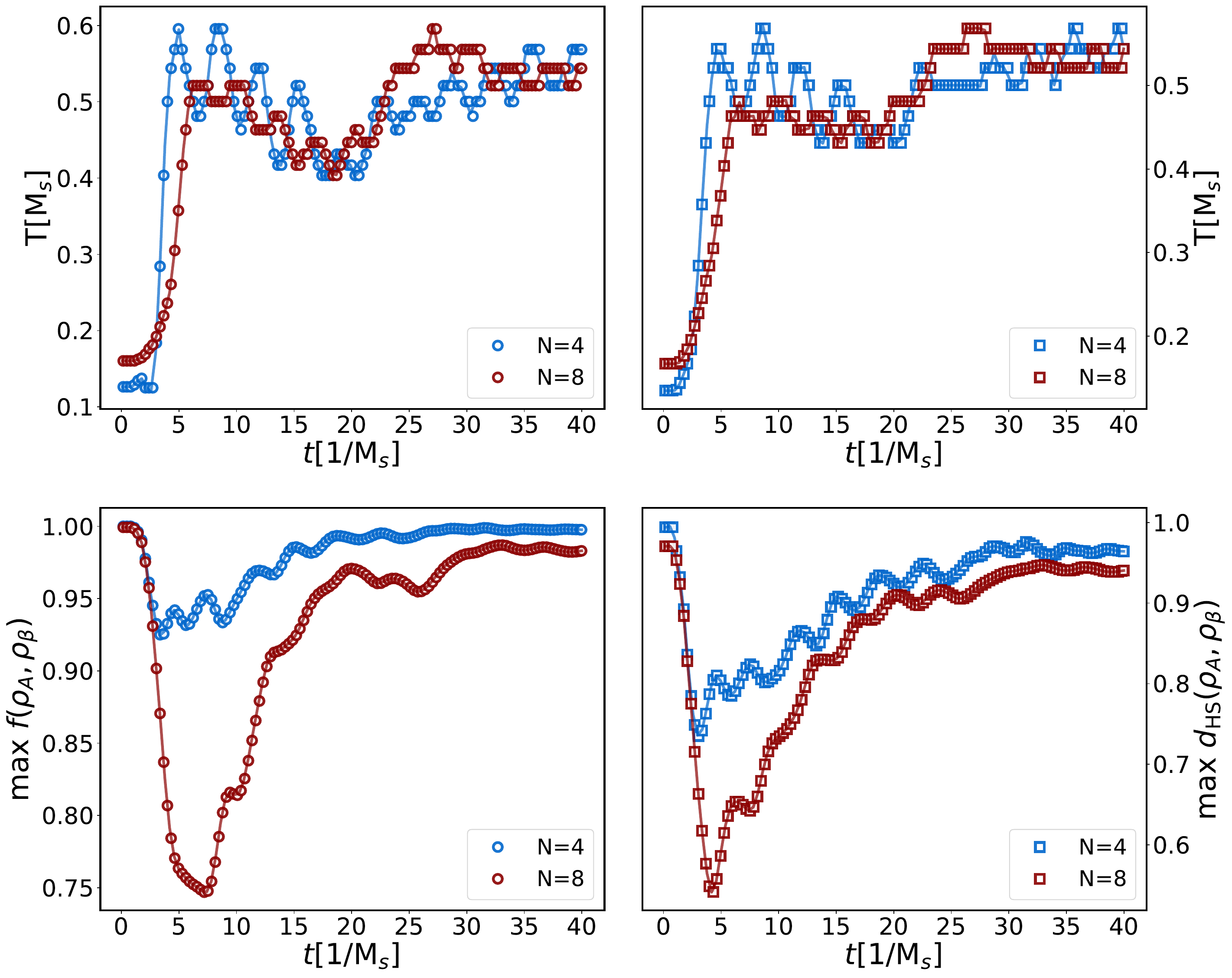}
    \caption{Top panel: The temperature as a function of time extracted from the maximal normalized overlap (right) and minimal HS distance (left) for subintervals $L=4,8$ centered in the middle. Bottom panel: the value of the metric at the extremum shown in the top panel. For the thermal states we used $N=12$.}
    \label{fig:overlaptime}
\end{figure}

Fig.~\ref{fig:overlaptime} shows the maximum value of overlap and complement HS distance as a function of time. We see that at early times the overlap is largest for thermal state with largest $\beta$ ($\beta=10$) since the system is still in vacuum. Upon switching on the sources agreement with thermal states drops significantly since we are far from equilibrium. Finally the two metrics start growing approaching one at late times. This suggest that $\rho_A$ becomes more and more similar to a thermal state as time passes, and the temperature can be extracted from the $\rho_\beta$ state that maximizes the metric. The average of the two temperatures at which the overlaps are maximal for $L=4,8$ is reported in Fig.~ \ref{fig:thermalization_m025}, labeled as ``overlap'', with the error corresponding to the standard error in such dataset. We find that it converges to the same value as temperature found from the other observables at late times.

\begin{figure}
    \centering
    \includegraphics[width=1\linewidth]{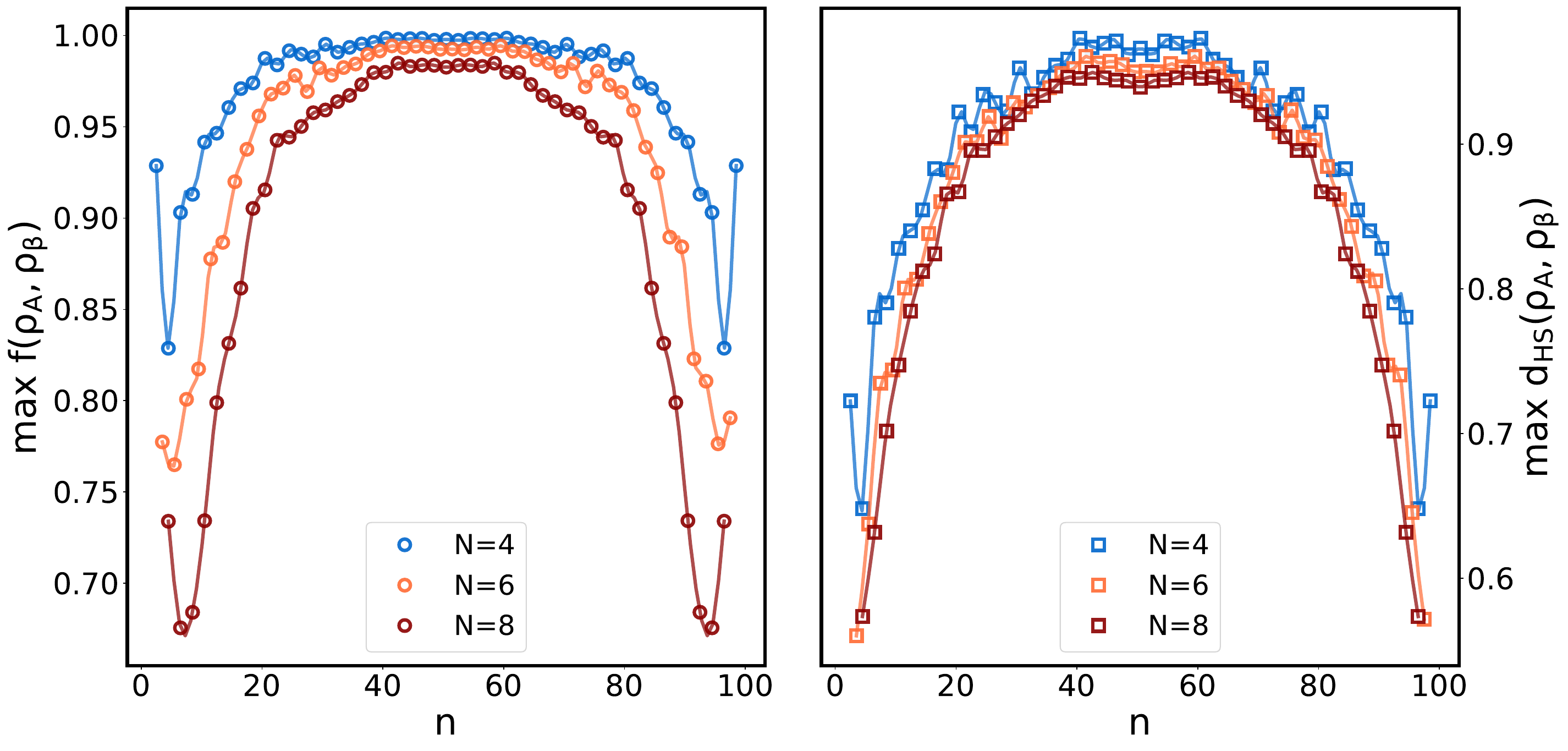}   
    \caption{Spatial cut at time $t\approx40/M_s$. We move a subinterval of size $N=4,6,8$ through the system. The $x$-axis indicates the center site of the subinterval. The left figure displays the maximal normalized overlap with the thermal states (obtained for $N=12$. The right plot shows the complement of the minimal HS distance.}
    \label{fig:overlapspace}
\end{figure}

We further investigate the local thermalization of the jet system. To do this, we take the final spatial configuration of the jet simulation and move the region $A$ from the left site of the lattice to the right, scanning the entire system. We also vary the size of the interval $A$. The value of our metrics are plotted in figure \ref{fig:overlapspace}. We can see that the maximal values of the metrics are always achieved in the central region of the lattice, whereas they tend to decrease toward the boundary. For the chosen time, the sources sit on the boundary sites and the region near them did not have time to thermalize yet. We note that the overlap remains close to maximal in a broader spatial region compared to the complement HS distance, which is due to the details of the metric functions employed. The fact that the two proxies do not exactly agree shows that thermalization is only partially achieved outside of the central region.

In a special case $L=2$, the reduced density matrix admits a simple decomposition on a complete basis of local operators with clear physical meaning, allowing us to connect the results about thermalization observed at the level of local operators and thermalization from the density matrix overlaps. These results are reported in App.~\ref{app:overlaps_L_2}.

\section{Hydrodynamic behavior} \label{sec:hydro}
This section addresses the time evolution of the jet system using hydrodynamics. This effective approach is commonly used to describe heavy-ion collisions and the Quark-Gluon Plasma formed therein, and has been remarkably successful, see for instance Refs. \cite{Jaiswal:2016hex,Florkowski:2017olj} for a review. An intense discussion is ongoing concerning the regime of applicability of relativistic hydrodynamics, as collective effects consistent with hydrodynamic behavior have been found in colliding systems composed of very few particles, for instance the ones created in collisions of proton-nucleus \cite{CMS:2012qk,STAR:2019zaf}, and even proton-proton \cite{CMS:2010ifv}, see \cite{Nagle:2018nvi,Schenke:2021mxx} for a review. The thermalization of the jet subsystem detailed in the previous sections offers the opportunity to compare the microscopic quantum dynamics with hydrodynamics and to test the effectiveness of the latter. Other studies about hydrodynamics from simulations of quantum systems can be found in Refs. \cite{Janik:2025bbz, Turro:2025sec}.

\begin{figure}
    \centering
    \includegraphics[width=0.9\linewidth]{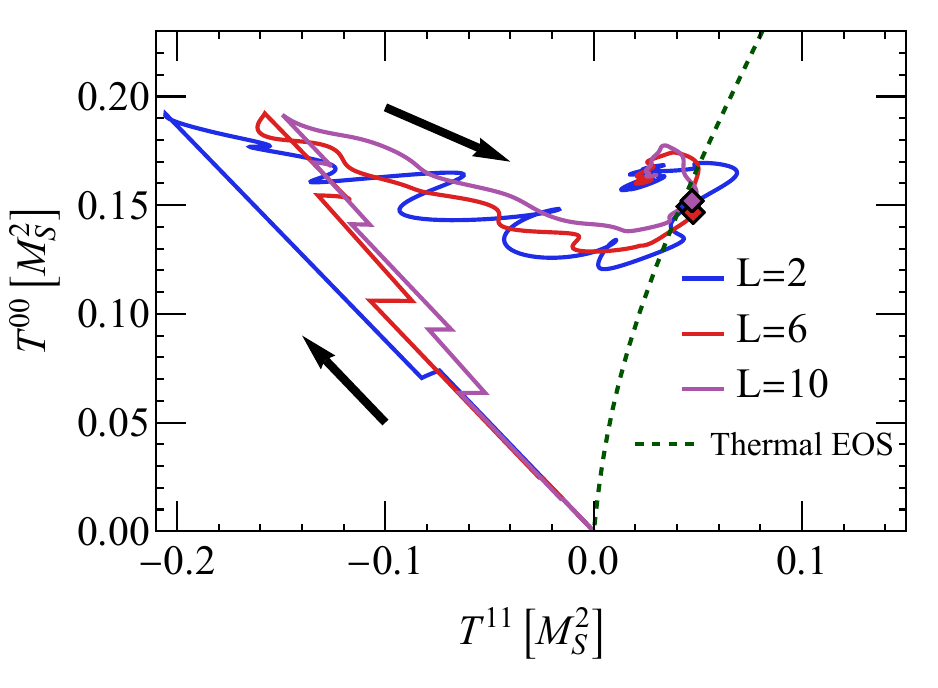}
    \caption{Energy density versus pressure modelled with $T^{00}$ and $T^{11}$, respectively, during the jet simulation. Three different subsystems with $L=2,6,10$ are shown. We subtracted the vacuum values; hence, the curves start in the origin ($t=0$) moving initially to negative pressures before turning around and approaching the thermal equation of state (EOS). The diamonds indicate $t_0\approx 21.4/M_s$. The arrows show the  flow of time.}
    \label{fig:equation of state}
\end{figure}

Throughout this section, we refer to the jet simulation results of the Schwinger model with $g=0.5/a$, $m=0.5g$ and $N=100$.  To compare quantum evolution with hydrodynamics, we first turn off the external sources at a time $t_0$, chosen such that the system is in a quasi-thermal state, but such that the jets are still relatively far from the boundaries of the lattice; in practice, we choose $t_0=21.4/M_s$. This step is motivated by the fact that the sources keep injecting energy in the system and it would be complicated to model them realistically in the hydrodynamic simulation; once they are turned off, energy and momentum are conserved, so hydrodynamics can be applied straightforwardly. We use the ideal fluid form for the energy momentum tensor, Eq.~\eqref{eq:ideal tmunu}, obeying the equations of motion $\partial_\mu T^{\mu\nu}=0$. To close the system of equations, the thermal equation of state(EOS) $P(T)$ is obtained with tensor networks methods, and we use standard thermodynamic relations to write equations of motion in terms of $T(t)$ and $u_x(t)$ only, i.e. the time-dependent temperature and the spatial component of the two-velocity, in Cartesian coordinates. Fig.~\ref{fig:equation of state} shows the equation of state obtained from the full simulation. Solid lines represent the evolution of energy density and pressure in time, which flows in the direction indicated by the arrows. The thermal equation of state is shown as a dashed line. We see that at small times, the pressure grows negative, similar to, for example, the empty MIT bag \cite{Johnson:1975zp,Khanmohamadi:2019jky}. The negative pressure may be associated with the presence of strings pulling fermions together, or with the inward pressure resulting from the partial destruction of the non-perturbative vacuum. Similar behavior of the pressure at early times was also obtained in holography \cite{Chesler:2008hg,Beuf_2009,Chesler:2009cy,Heller_2012,PhysRevD.85.126002}. After a while, the string breaks forming fermion-antifermion pairs, which increase the pressure. Eventually, this process brings the equation of state closer to the thermal expectation. The time $t_0$ when we initialize the hydrodynamic evolution is shown in Fig. \ref{fig:equation of state} with highlighted markers, and we can see that at that stage the equation of state is close to the thermal one. We notice that these results are somewhat different from those obtained in Ref.~\cite{Janik:2025bbz}.

The equations of motion are solved using Fluid$u$M \cite{Floerchinger:2018pje,Devetak:2019lsk,Capellino:2023cxe}, an accurate numerical scheme developed for the phenomenology of heavy ions. The spatial grid is chosen to reproduce the physical lattice of the tensor network simulation, and the temperature is initialized from the energy density profile at the switch-off time. The initial two-velocity is $u^\mu = (1,0)$: we will always consider the center of the expanding jet system, which is at rest in good approximation. 

\begin{figure}
    \centering
    \includegraphics[width=1\linewidth]{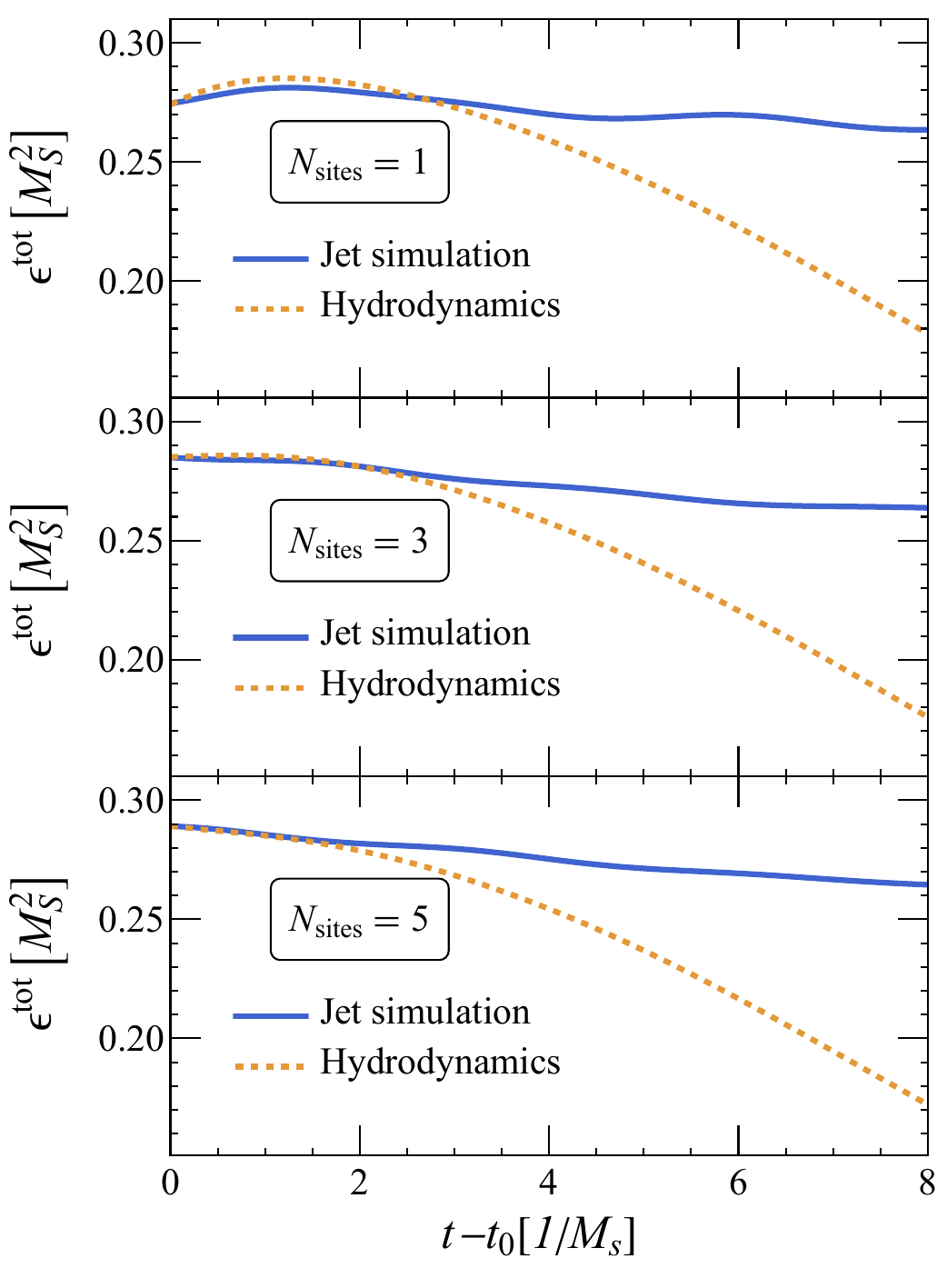}
    \caption{Energy density as a function of time for the central physical region corresponding to $N_{\rm sites}=1$, $3$ and $5$ sites. Solid blue lines correspond to the jet simulation results, dashed orange lines are the results from Fluid$u$M.}
    \label{fig:hydro}
\end{figure}

Our results are reported in Figure~\ref{fig:hydro}. We compute the energy density for different sizes of the jet subsystem, corresponding to $N_{\rm sites}=1$, $3$, and $5$ central sites. The solid line is the result of the jet simulation, whereas the dashed lines represent the results from Fluid$u$M.

We see that there is a rather good agreement between Fluid$u$M and the jet simulation for some initial time, until eventually the energy density of the jet simulation plateaus and the hydrodynamic one decays to zero. This difference allows identifying a sort of ``freeze-out time'' of the intra-jet medium, when the energy is effectively converted to final-stage bound states. Indeed, the jet simulation takes into account particle production and reaching the plateau in the energy density represents the energy of the bound states produced during the evolution. It cannot be reproduced by hydrodynamics describing the system expansion in vacuum, where the medium is bound to become, eventually, infinitely diluted. Although the simulation at hand does not represent a heavy-ion collision, the observation of a freeze-out in the intra-jet medium is remarkable. 

This preliminary study shows that an hydrodynamic description of quantum dynamics is possible and qualitatively accurate up to the freeze-out stage. The initialization time plays a crucial role, similar to what is observed in heavy-ion simulations \cite{Song:2008si}.
From this preliminary study, different future directions are possible. For instance, it would be interesting to study the effects of different couplings and masses, the generation of sound and dissipative modes and the explicit calculation of transport coefficients, such as bulk viscosity, from real time correlators. These aspects will be addressed in a future publication.

\section{Conclusion} \label{sec:conclusion}
The results presented above show how apparently thermodynamical behavior in a driven closed quantum system emerges in real time. Our setup mimics the production of jets in high-energy physics. Although conventional mechanisms of thermalization (due to rescattering of produced particles) are very unlikely to yield thermalization in this process, experimental observations point to thermal abundances of the produced hadrons. In general, ``early thermalization" is a salient feature of high-energy inelastic hadron and heavy-ion collisions. Here, we argue that this apparent thermalization is a direct consequence of quantum entanglement.
\vskip0.3cm

We find that the expectation values of local operators as functions of time approach their thermal counterparts. Moreover, the overlap between the evolving density matrix and the thermal one becomes very large at late times. The dynamics of the energy-momentum tensor at intermediate stage of the evolution (after thermalization but before the ``freeze-out", when the meson states are formed) is well described by relativistic hydrodynamics. 
\vskip0.3cm
Our results elucidate the mechanisms by which quantum entanglement drives thermalization in closed field-theoretic systems, and how the thermodynamic behavior emerges from unitary quantum dynamics.
The approach to maximal entanglement and thermalization in the fragmentation of jets motivates new venues for experimental studies that have already begun, with very promising results \cite{Datta:2024hpn}.
\vskip0.3cm

\section*{Acknowledgment}

We are grateful to the Simons Center for Geometry and Physics, Stony Brook University for organizing the workshop ``Entanglement, thermalization, and holography" that has inspired the initial stages of this work. We are grateful to Anatoly Dymarsky and Michal Heller for useful discussions. AF acknowledges particularly useful discussions on MPS with Andrea Bulgarelli, Stefano Carignano, Olivier Gauth\'e, Luca Tagliacozzo and Andreas Weichselbaum. AP is grateful to Eduardo Grossi for discussions about Fluid$u$m. This work was supported by the U.S. Department of Energy, Office of Science, Office of Nuclear Physics, Grants No. DE-FG88ER41450 (D.K., A.P.) and DE-SC0012704 (A.F., S.G., D.K.) and by the U.S. Department of Energy, Office of Science, National Quantum Information Science Research Centers, Co-design Center for Quantum Advantage (C2QA) under Contract No.DE-SC0012704 (A.F., S.G., D.K.). A.F. is supported by the DFG through Emmy Noether Programme (project number 545261797. D.F. was supported by DOE under contract No. DE-SC0012704 and by BNL LDRD project No. 25-033B. S.G. was supported in part by a Feodor Lynen Research fellowship of
the Alexander von Humboldt foundation.
S.S. is supported by National Key Research and Development Program of China under Contract No. 2024YFA1610700 and by Tsinghua University under grants No. 043-531205006 and No. 043-53330500.

\clearpage
\appendix 

\section{Spin basis}
\label{app:spin}

There is a direct mapping of fermionic degrees of freedom onto qubits in 1+1 dimensions. It is given in  terms of Jordan-Wigner transformation
\begin{equation}
     \chi_n =\frac{X_n-iY_n}{2}\prod_{j=1}^{n-1}(-i Z_j)
\end{equation}
In this formulation the Hilbert space is $(\mathbb{C}^2)^{\otimes N}$, arranged in a 1-dimensional lattice with spin 1/2 at each of $N$ sites. $X_i, Y_i$ and $Z_i$ are Pauli operators acting on the $i$-th site. With such transformation the kinetic and mass terms of Eq.(\ref{eq:lat_Ham}) become
\begin{align}
    H_{kin} &= \frac{1}{4a}\sum_{n=1}^{N-1}X_n X_{n+1} + Y_n Y_{n+1} \, , \\
    H_m &= \frac{m}{2}\sum_{n=1}^{N} (-1)^n Z_n \, ,
\end{align}
and the electric charge operator is $Q_ i = \dfrac{Z_i + (-1)^i}{2}$. The local operators comprising the set $\cal O$ in Eq. (\ref{eq:O_set}) are expressed in the spin representation as 
\begin{align}
\widehat{\bar\psi\psi}_n &= (-1)^n \frac{Z_n}{2}\, , \\
    \hat K_{n} &= \frac{X_n X_{n+1} + Y_n Y_{n+1}}{4}\, .
\end{align}

With an eye towards a future implementation on a quantum computer, we perform some of our simulations directly in the spin basis. Namely, we choose to do all the finite temperature calculations, either with exact diagonalization or with tensor networks, in the spin basis. Of course, the results do not depend on whether one uses fermion or spin representation.

\section{Thermal expectation values} \label{sec:thermal_prelim}
\subsection{Exact diagonalization}
Thermal expectation values of local operators can be obtained in the lattice Schwinger model using different Hamiltonian approaches. For sufficiently small systems, the exact spectrum of the Hamiltonian can be obtained using exact diagonalization:
\beq
H |E_n\rangle = E_n |E_n\rangle \, .
\eeq
Note that here we discuss thermal properties of the Schwinger model without external sources, and the corresponding Hamiltonian in the charge 0 sector is given by Eq.(\ref{eq:lat_Ham}) at $t=0$. For nonzero total charge Gauss's law should be resolved differently in order to preserve $\cal CP$ symmetry. We refer the reader to Appendix \ref{app:L_charge_Q} for the formulation of the electric field Hamiltonian used in a general charge sector.

A thermal expectation value of an operator $\cal O$ is constructed by summing up contributions from all the eigenstates with the Boltzmann weights:
\beq \label{eq:boltzmann}
\langle {\cal{O}}\rangle_T  = \frac{\sum_n e^{-E_n/T} \langle E_n|{\cal{O}}|E_n\rangle}{\sum_n e^{-E_n/T}} = \Tr(\rho_T {\cal O}) \, ,
\eeq
where the Boltzmann constant is set to unity and in the last equality we defined the thermal density matrix $\rho_T = \dfrac{e^{-\hat H/T}}{\Tr\, e^{-\hat H/T}}$. The $U(1)$ symmetry of the Hamiltonian can be leveraged to reduce the computational cost by working in fixed charge sectors. If $Q = \sum_{n=1}^N Q_n$ is the  operator of the total system's charge, the Hamiltonian is block-diagonal in the fermionic occupation basis with each block corresponding to a different value of $\langle Q\rangle = Q^*$. Thus, one can work in the Hilbert space of dimension $\begin{pmatrix} N/2-Q^* \\N
\end{pmatrix}$ in the corresponding charge sector instead of dealing with the full Hilbert space of dimension $2^N$. We will focus on the grand canonical ensemble  with the chemical potential $\mu = 0$, that amounts to summing over all the charge sectors. The reason is that even though the full system with two jets stays charge neutral during the entire evolution, we will be interested in the subsystem located at the center, where the charge can fluctuate.

Even using the block decomposition of the Hamiltonian the exact diagonalization method is computationally demanding. Due to hardware limitations we only use it for $N$ up to 14. As the correlation length is of order of few lattice steps one may worry that boundary and finite volume effects affect the observables in the bulk of the system with such a limited system size. We confirm this at this end of this appendix, and refer the impatient reader to Fig.~\ref{fig:local_obs_vs_T}.

With exact diagonalization one can also find the temperature dependence of the Gibbs entropy in a straightforward way:
\begin{align}
    S(T) &= -\sum_n p_n(T)\log p_n(T) \label{eq:S(T)} \\
\end{align}
where $p_n(T) = e^{-E_n/T}/Z$ are the Boltzmann weights. 

\begin{figure}
    \centering
\includegraphics[width=\linewidth]{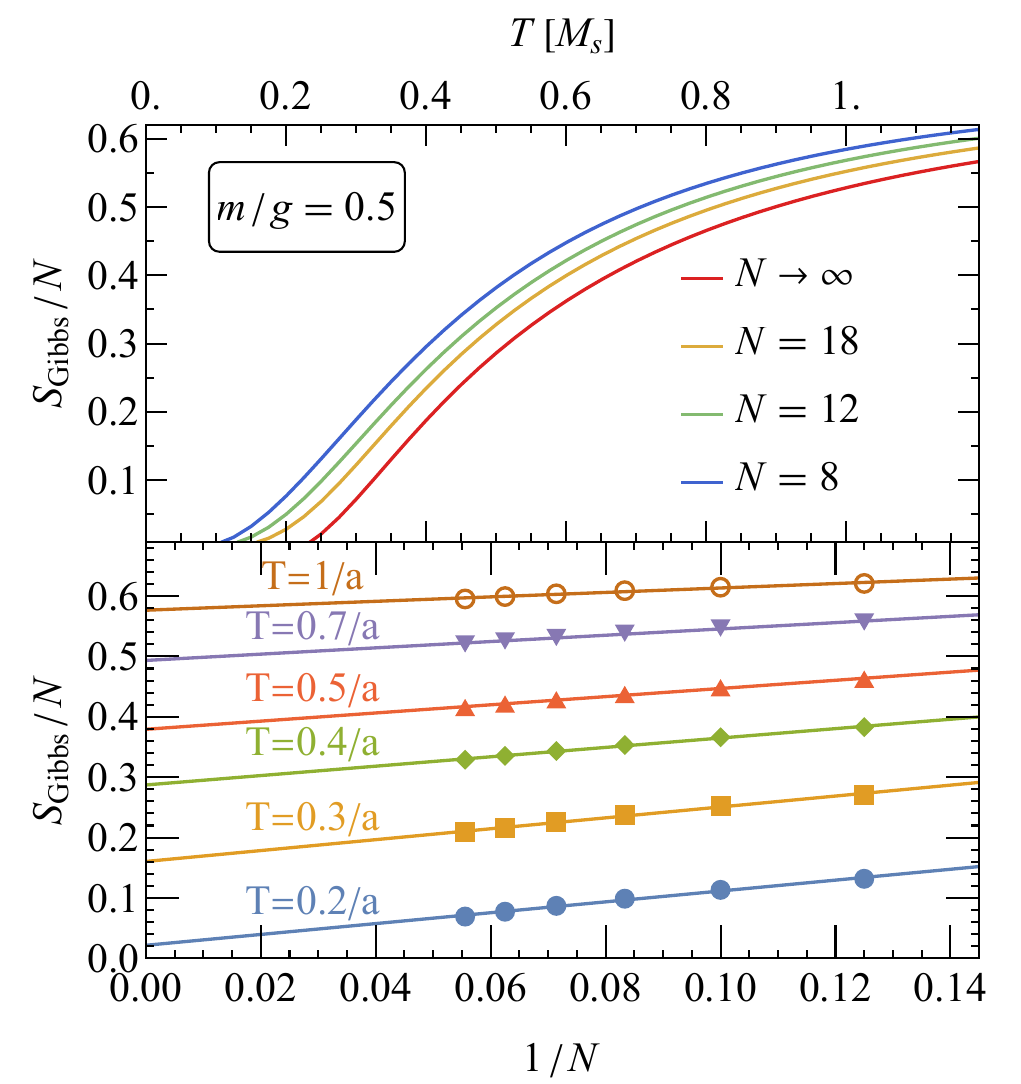}
    \caption{Top panel: Gibbs entropy density  for different values of the system size $N$ as a function of temperature obtained with exact diagonalization as explained in the main text. Bottom panel: linear extrapolation in $1/N$ of the Gibbs entropy density for several values of the temperature. The intercepts with the $y$ axis correspond to the infinite volume extrapolation and are used in the top panel.  Fermion coupling and mass are $g=0.5/a, m=0.25/a$. }
    \label{fig:S_vs_N}
\end{figure}

For the purposes of finding the full spectrum of the Hamiltonian the largest volume we managed to access within our resources is $N=18$. It is thus important to consider a thermodynamics extrapolation. We consider several values of the system volume in the range $N\in[8,18]$ and perform a linear extrapolation of the Gibbs entropy density in $\frac{1}{N}$, shown in the bottom panel of Fig. \ref{fig:S_vs_N}. Similar to finite volume effects for the local observables, we find that they are the strongest at low temperatures. The top panel of Fig. \ref{fig:S_vs_N} displays the temperature dependence of the Gibbs entropy for different system sizes as well as its infinite volume extrapolation. For the purpose of extracting the effective temperature we use the function $T_S$ inverse to the infinite volume extrapolated $S(T)$. 

Note that while all observables are spatially uniform,  finite lattice spacing and open-boundary conditions break translation invariance. In all our thermal computations, we mitigate these effects by considering $l/2=3$ physical lattice sites at the center of the lattice and performing an average. 

\subsection{Purification and MPS} \label{app:purification}

In order to address the finite volume issue we perform finite temperature tensor network simulations for larger system sizes. We use the purification method to extract thermal expectation values of local observables \cite{Verstraete:2004gdw, Zwolak:2004nwu,Feiguin:2005jud}.
A thermal system is described in terms of a density matrix rather than a state vector. In the tensor network language it is represented by an MPO rather than an MPS. However, there is a mapping between state vectors in a $D^2$-dimensional Hilbert space and operators, particularly density matrices, in a $D$-dimensional Hilbert space. The idea of the purification method is to represent a thermal density matrix with a state vector in an enlarged Hilbert space. All manipulations, such as state preparation, (imaginary or real) time evolution and computing observables, can be performed at the level of the state vector with well-established MPS techniques. The density matrix of a thermal state can be recovered as a reduced density matrix corresponding to the original Hilbert space.

Let us focus on a spin system for clarity. In order to describe a thermal state in Hilbert space $\cal H$ of $N$ qubits, one considers an enlarged Hilbert space $\tilde {\cal H} = {\cal H}\otimes{\cal H'}$ where $\cal H'$ has the same dimension $2^N$ as $\cal H$ and is called the ancilla space. The product state of Bell states 
\begin{equation}
    |\tilde\Psi(0)\rangle  = \bigotimes_{i=1}^N \frac{|0\rangle_i|0\rangle_i' + |1\rangle_i|1\rangle_i'}{\sqrt{2}}  
\end{equation}
yields a maximally entangled state when traced over $\cal H'$ and thus is a purified description of a density matrix at  the inverse temperature $\beta = 0$. By construction, it is naturally represented as an MPS of bond dimension 1 and thus the MPS representation is very efficient. To obtain a state describing the system at finite temperature, an imaginary time evolution is necessary:
\begin{equation}
    |\tilde \Psi(\beta)\rangle = \frac{1}{\sqrt{Z(\beta)}}e^{-\beta \hat{\tilde H}/2} |\tilde \Psi(0)\rangle \,,
\end{equation}
where the Hamiltonian in the enlarged Hilbert space acts nontrivially only on the physical space: 
\begin{equation}
    \hat{\tilde H}_{\tilde{\cal H}} \equiv  \hat H_{{\cal H}} \otimes I_{\cal H'}\,,
\end{equation}
and 
\begin{equation}
    Z(\beta) \equiv \langle\tilde\Psi(0)|e^{-\beta \hat{\tilde H}}|\tilde\Psi(0)\rangle\,.
\end{equation}
This imaginary time evolution can be performed with a variety of methods designed for MPS, in particular in this work we use TDVP. With an MPS representation of a thermal state the expectation values of local operators are found using a contraction with an MPO representation of the corresponding operator in the enlarged Hilbert space:
\begin{equation}
    \langle{\cal O}\rangle_\beta = \langle\tilde\Psi(\beta)|{\cal O_H}\otimes I_{\cal H'}|\tilde\Psi(\beta)\rangle\,.
\end{equation}

\begin{figure}
    \centering
    \includegraphics[width=\linewidth]{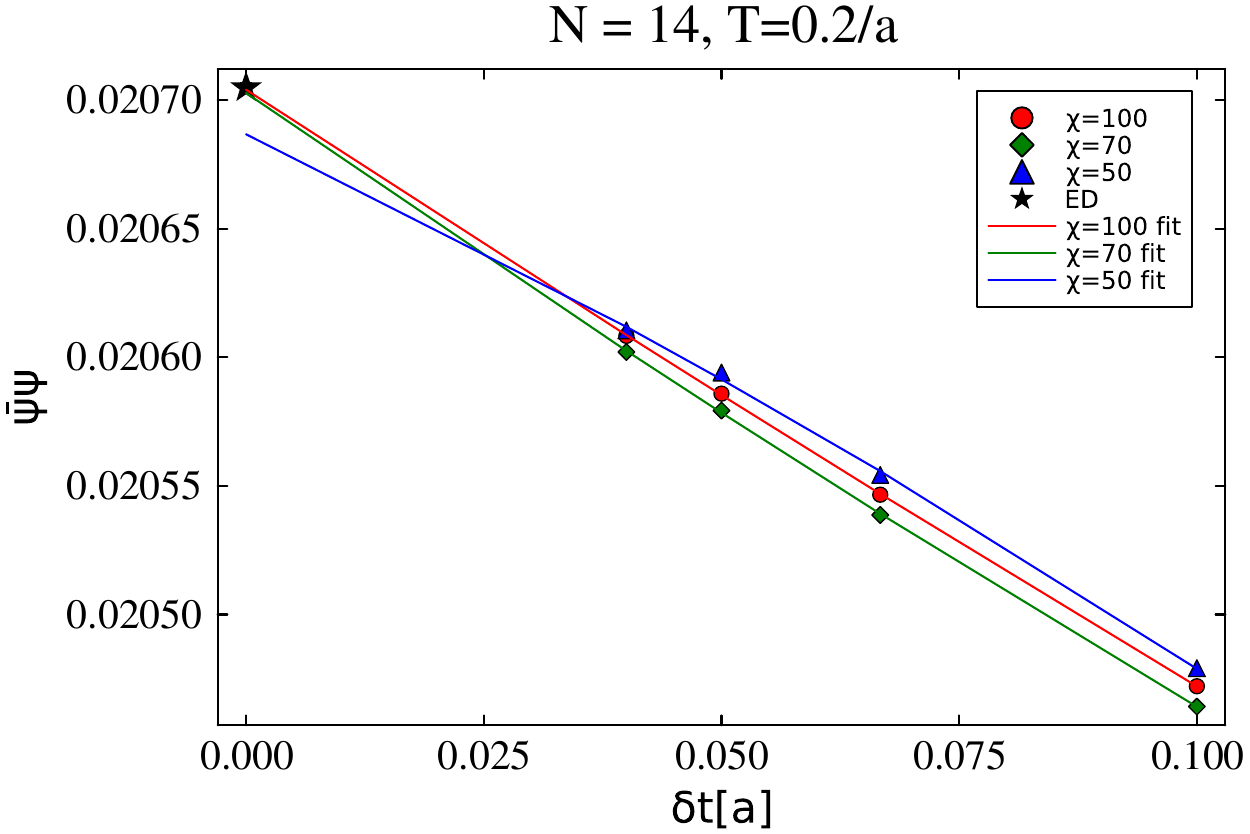}
    \caption{
    Comparing purification to exact diagonalization (ED) using chiral condensate $\bar\psi\psi$ (with the vacuum value subtracted) as an observable. Purification results for various fixed bond dimension values $\chi$ are shown against the imaginary time evolution timestep $\delta t$. Quadratic fits in $\delta t$  are performed to extrapolate to $\delta t\rightarrow0$. The extrapolation greatly improves the agreement with the exact result. Simulations with the bond dimension larger than 100 coincide with the one as 100, so at $\chi=100$ the MPS truncation error is already negligible.}
    \label{fig:N-14_trotter_step_extrapolation}
\end{figure}

As one of the checks that the purification algorithm is working correctly we analyze how well it matches against exact diagonalization. With the latter the largest system size that we simulated is $N=14$. Access to the exact result allows us to examine systematic uncertainties of purification and study how they can be reduced.

Purification algorithm possesses systematic errors typical for any tensor network method involving time evolution: error from a finite timestep and error from truncating the MPS at  a given bond dimension (or at a given SVD eigenvalue). Other uncertainties, such as finite volume and finite lattice spacing effects, will not be of interest to us in this subsection as we are comparing two descriptions of the same lattice system with a fixed volume and lattice step.

Finite TDVP timestep errors include Trotter errors that grow with the timestep size $\delta t$ and projection errors that grow with the number of timesteps. Therefore, for the optimal result one needs to maintain a balance between the two errors and take $\delta t$ that is neither too large nor too small. For this reason, we keep $\delta t$ finite, probing several different values, and find the range of  $\delta t$ such that an observable can be efficiently fitted using a 2nd order polynomial:
\begin{equation}
    f(\delta t) = f_0 + f_1 \delta t + f_2 \delta t^2 ,
\end{equation}
Here $f$ can stand for any observable; as explained in the main text we focus on the chiral condensate, kinetic energy density and the condensate 2-point function. $f_0$ represents the $\delta t\rightarrow 0$ extrapolation of the observable.

This procedure is demonstrated in Fig. \ref{fig:N-14_trotter_step_extrapolation} for chiral condensate at temperature $T=0.2/a$, where we also present a comparison to the exact diagonalization and investigate the effect of varying the bond dimension of the MPS. In the main text, we always use such values of the bond dimension that ther results are converged. As one can see, even before the Trotter step extrapolation purification produces results in very good agreement with the exact diagonalization results. Trotter step extrapolation allows to further improve this agreement to a relative error $\sim 10^{-4}$.

\begin{figure}
    \centering
    \includegraphics[width=\linewidth]{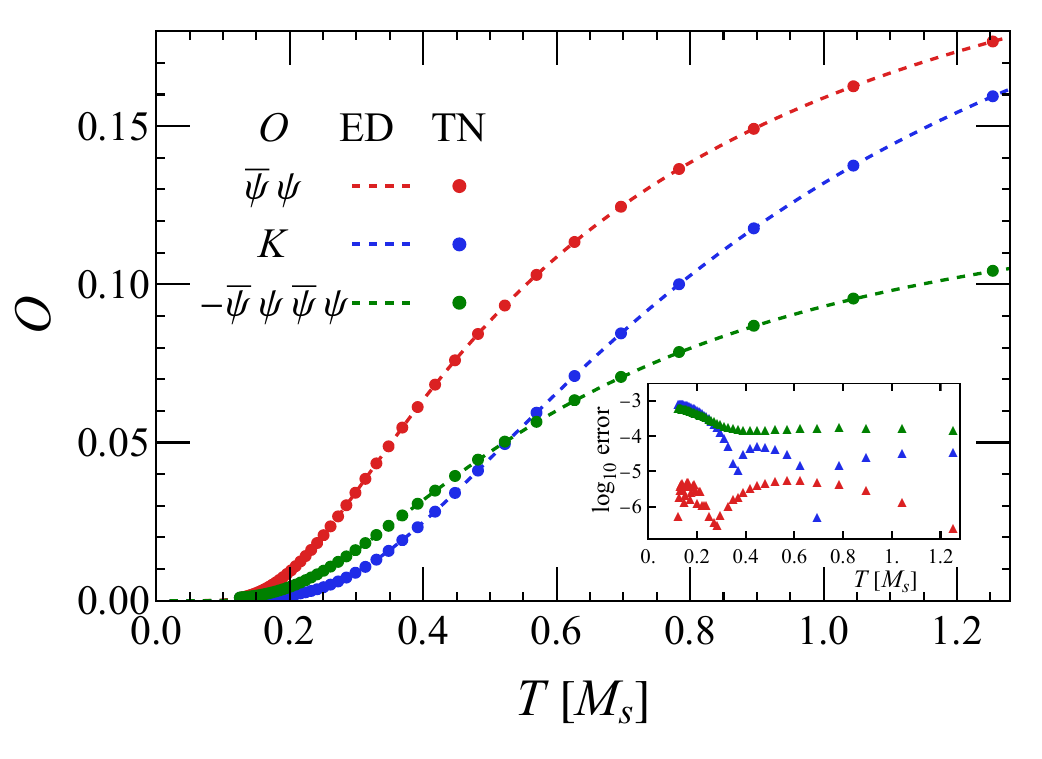}
    \caption{Local observables (chiral condensate $\bar\psi\psi$, kinetic energy density $K$ and condensate-condensate correlation function $\bar\psi\psi\bar\psi\psi$) obtained with exact diagonalization (ED)  and the purification algorithm realized with tensor network methods (TN) shown as a function of temperature. Inset: logarithm of the absolute value of the difference between ED and TN.}
    \label{fig:TN_ED}
\end{figure}

In Fig. \ref{fig:TN_ED} we present evidence of the agreement between ED and TN, improved by the Trotter step extrapolation, extending across the set of local observables and the range of temperatures that we study. It confirms the validity of the purification method used throughout this study.

\begin{figure}
    \centering
\includegraphics[width=0.75\linewidth]{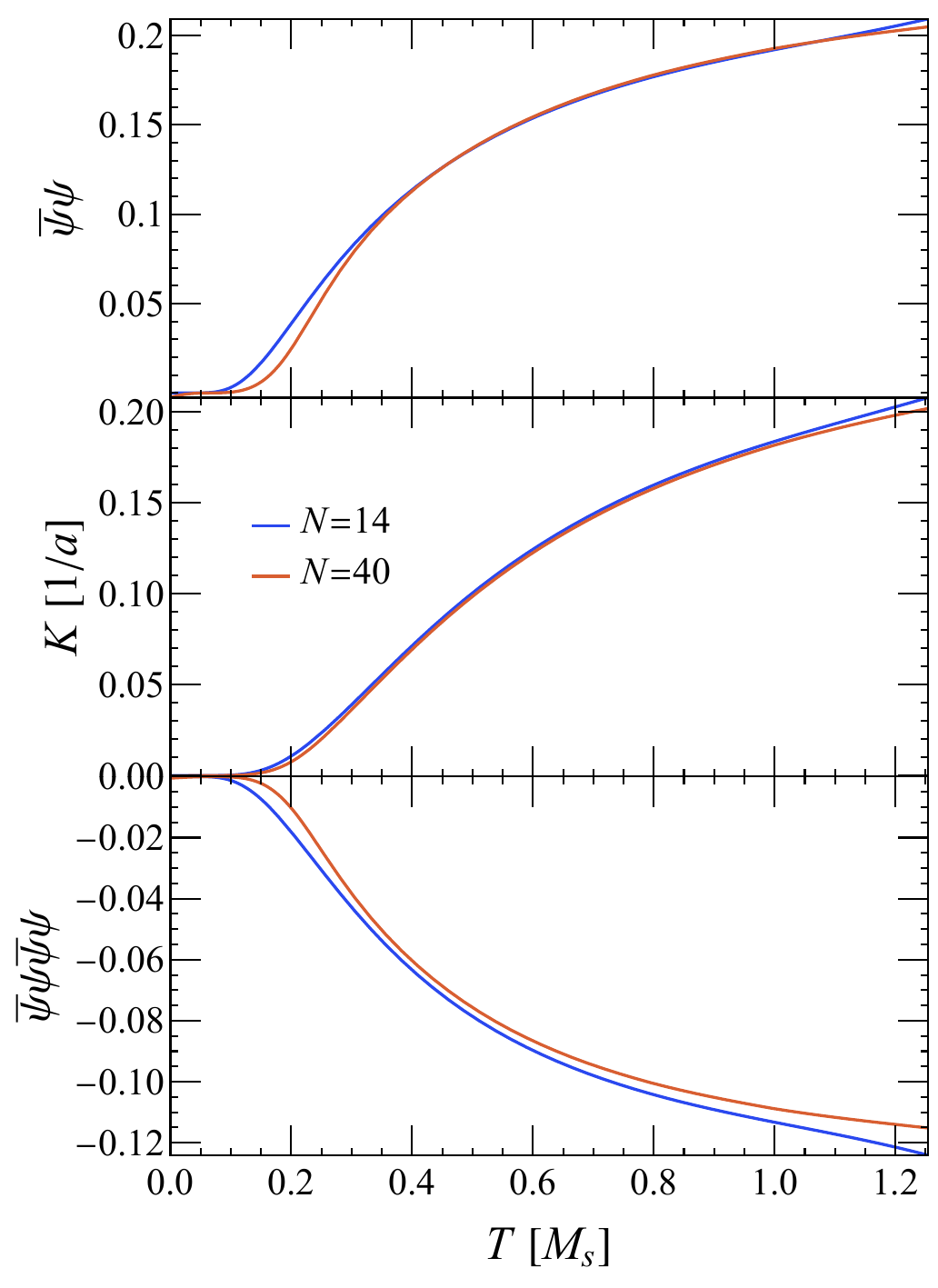}
    \caption{Average values of local observables (top panel: chiral condensate, middle panel: kinetic energy, bottom panel: condensate 2-point function) at the center of the system as a function of temperature for two different values of the system size $N$. $N=14$ is the largest system size studied with exact diagonalization. $N=40$ results are calculated using purification algorithm with tensor network methods. The vacuum expectation values value is subtracted. The fermion charge and mass are $g=0.5/a, m=0.25/a$. Temperature is expressed in the units of the pseudoscalar meson mass, see Appendix \ref{app:meson} for more details. } 
    \label{fig:local_obs_vs_T}
\end{figure}

Finally, we confirm that the size of the finite volume correction is not too large. We show in Fig. \ref{fig:local_obs_vs_T} a comparison between results obtained on a lattice of size $N=14$ and $N=40$. The disparity between the results is found to be moderate, thus the finite volume effects can be assumed small at $N=40$.

\section{Pseudoscalar meson mass} \label{app:meson}

In the main text dimensionful quantities are expressed in terms of a physical (pseudo)scalar mass $M_S$ in order to facilitate the comparison between systems with different values of the fermion mass $m$. The scalar ``meson" with zero momentum is the first excitation in the spectrum of lattice Schwinger model, thus its mass can be readily extracted as the energy of the first excited state~\cite{Grieninger:2024cdl,Grieninger:2024axp}. In a lattice system it receives small finite volume corrections. As we work with a variety of system sizes, we found it most convenient to normalize the results with respect to the infinite volume extrapolation of the scalar mass.

\begin{figure}
    \centering
    \includegraphics[width=1\linewidth]{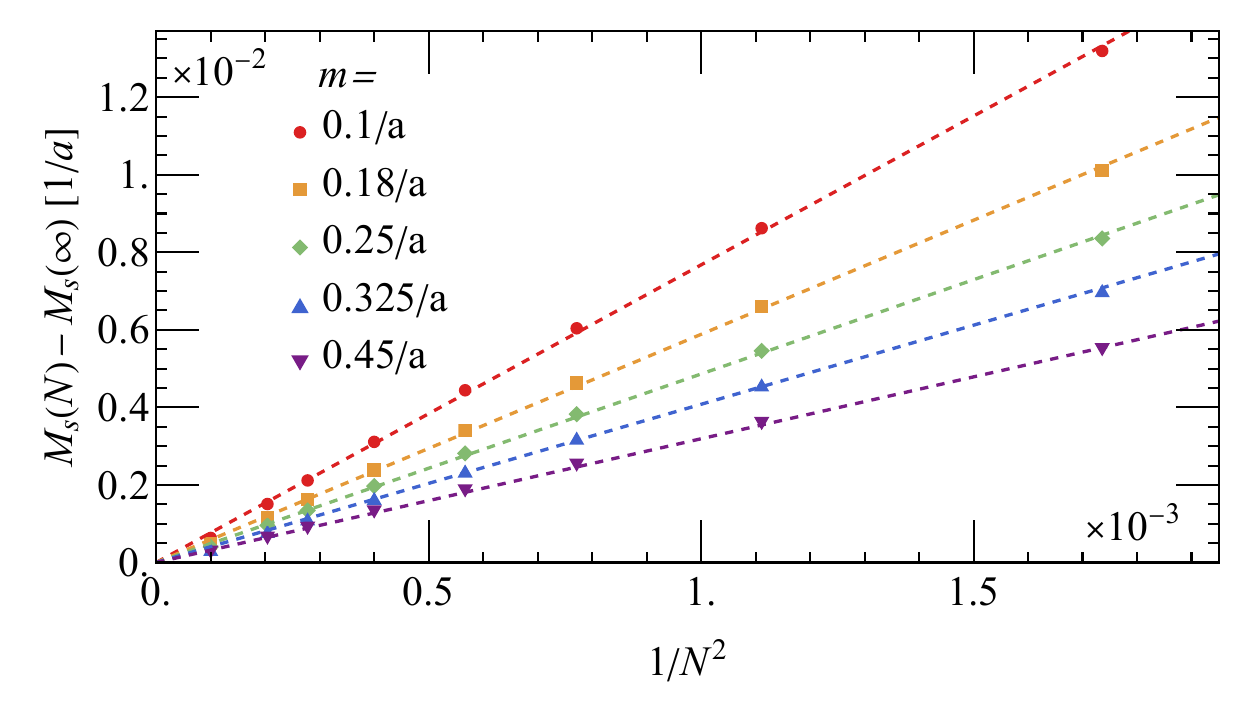}
    \caption{Masses of the first excited states shown for a variety of fermion mass values $m$ as functions of the inverse square of the system size $N$. Linear fits for infinite volume extrapolation are shown in dashed lines. Infinite volume extrapolated values are subtracted. Fermion coupling is $g=0.5/a$.}
    \label{fig:meson_mass}
\end{figure}

We have found the energies of the ground state and the first excited state for several different values of the system size $N$ using DMRG. For the infinite volume extrapolation we use the fit \cite{Banuls:2013jaa}
\begin{equation}
    M_S(N) = M_S(\infty) + \frac{\alpha}{N^2} + O\big(\frac{1}{N^3}\big)\, .
\end{equation}
As can be seen from the results, displayed in Fig. \ref{fig:meson_mass}, this fit describes the data very well and thus we obtain a reliable estimate of the infinite volume limit of the scalar mass, presented in Tab. \ref{tab:meson_mass}. Note that we do not attempt to do a continuum extrapolation as throughout the paper we work exclusively with the lattice version with the same value of lattice spacing as the one used to find the meson mass. 

\begin{table}
    \centering
    \begin{tabular}{|c|c|c|c|c|c|}
    \hline
        $m\,[1/a]$ & 0.1 & 0.18 & 0.25 & 0.325 & 0.45 \\
        \hline
        $M_s\,[1/a]$ & 0.518  & 0.667 & 0.798 & 0.939 & 1.177 \\
        \hline
    \end{tabular}
    \caption{Scalar masses $M_s$ for a variety of fermion mass values $m$. Infinite volume extrapolation is performed, see Fig.~\ref{fig:meson_mass} and discussion in the text. Fermion coupling is $g=0.5/a$.}
    \label{tab:meson_mass}
\end{table}

\section{Entanglement spectrum for the center of the chain}
\label{app:entanglement_spectrum}

\begin{figure}
\includegraphics{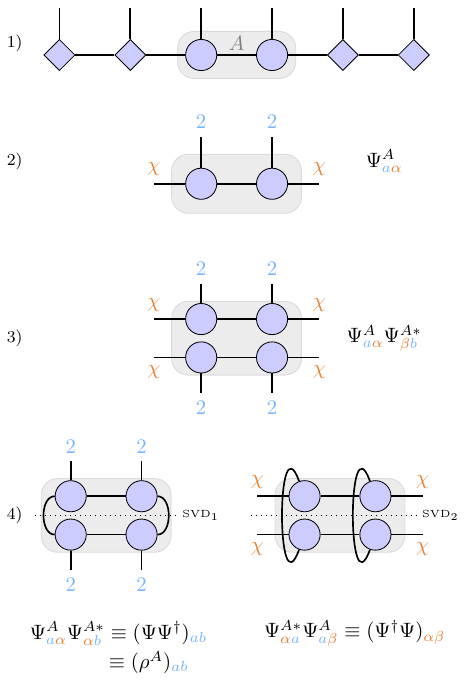}
\caption{Procedure to compute the entanglement spectrum for a central region of the spin chain. Diamonds in the first line illustrate the fact that the center of orthogonality is chosen within $A$. Latin indices refer to the ``physical'' dimension of $A$ and run from $1$ to $2^{l}$ for $A$ comprising $l$ sites. Greek indices run over the ``virtual'' space and take values between $1$ and the bond dimension $\chi^2$.}
\label{fig:MidEntropyMPS}
\end{figure}

The entanglement spectrum of a subregion $A$ of size $l$ in the center of the lattice can be obtained with a numerical effort that scales like $\chi^6$, with $\chi$ the bond dimension\footnote{We thank S.~Carignano, for pointing this out to us and A.~Bulgarelli and L.~Tagliacozzo for further explanations.}. We illustrate this procedure in \cref{fig:MidEntropyMPS} using tensor network diagrams.

\begin{enumerate}
    \item  Select a subregion $A$ and choose the center of orthogonality of the MPS to lie within this subregion. This achieves that all the tensors to the left/right of $A$ are in left/right canonical form; this is illustrated by drawing diamonds for the tensors outside of $A$. 
    \item Drop the tensors outside of $A$. It will be useful to think about the resulting object $\Psi_{a\alpha}$ as a map between the physical Hilbert space $\mathcal{H}^A, \mathrm{dim}(\mathcal{H}^A)=2^{l}$ and the ``virtual space'' of dimension $\mathcal{V}, \mathrm{dim}(\mathcal{V})=\chi^2$. Latin letters are used to denote physical indices and Greek letters are used for virtual indices.
    \item Construct the outer product $\Psi^A_{a\alpha}\Psi^{A*}_{\beta b}$.
    \item An MPO representation of the reduced density matrix is shown on the left-hand side of the last row of \cref{fig:MidEntropyMPS}. It is simply obtained by tracing over the virtual indices. One can in principle obtain the spectrum from this object by performing a SVD (SVD$_1$), but this scales exponentially in the system size $O(2^{3l})$. Fortunately, the object drawn on the right-hand side of the row, obtained by contracting the $physical$ indices, i.e. $\Psi^{A*}_{\alpha a}\Psi^{A}_{a\beta}$, has the same non-zero singular values. Computing the SVD of this object only scales as $O(\chi^6)$.
\end{enumerate}

\section{Symmetry-resolved entanglement entropy and thermalization by charge sector} \label{app:symmetry_resolved}

Due to charge conservation in the Schwinger model each eigenstate of the reduced density matrix $\rho_A$ has a definite charge $Q$. Therefore, the entanglement spectrum can be resolved by symmetry sectors: 
\begin{equation}
    \{\lambda_i\} = \left\{\bigcup_{Q = -L/2}^{L/2} \lambda_i^Q \right\}.
\end{equation}

\begin{figure}
    \centering
    \includegraphics[width=0.45\textwidth]{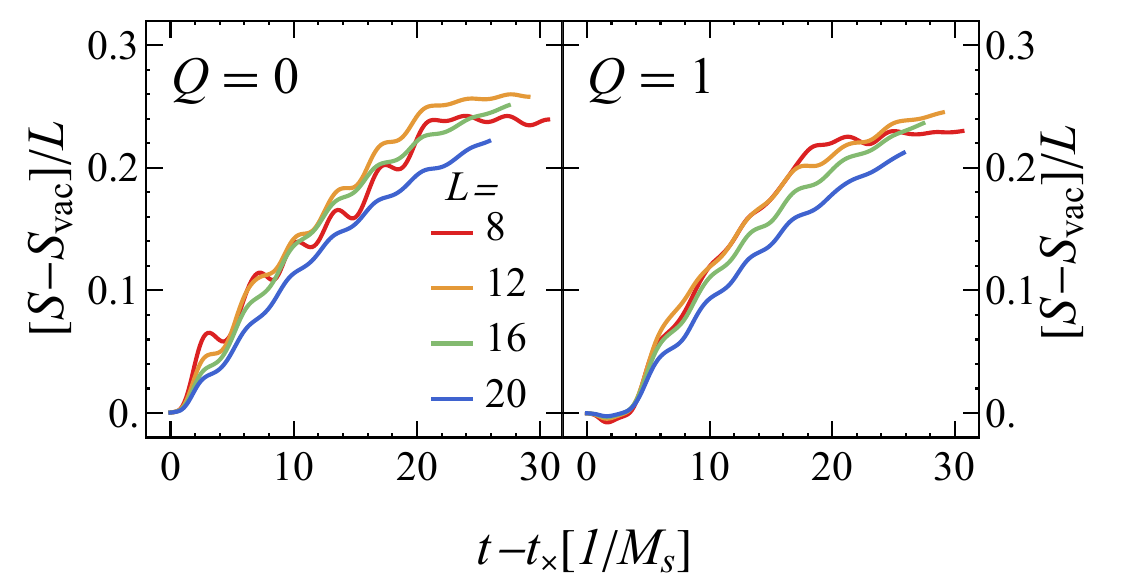}
    \includegraphics[width=0.45\textwidth]{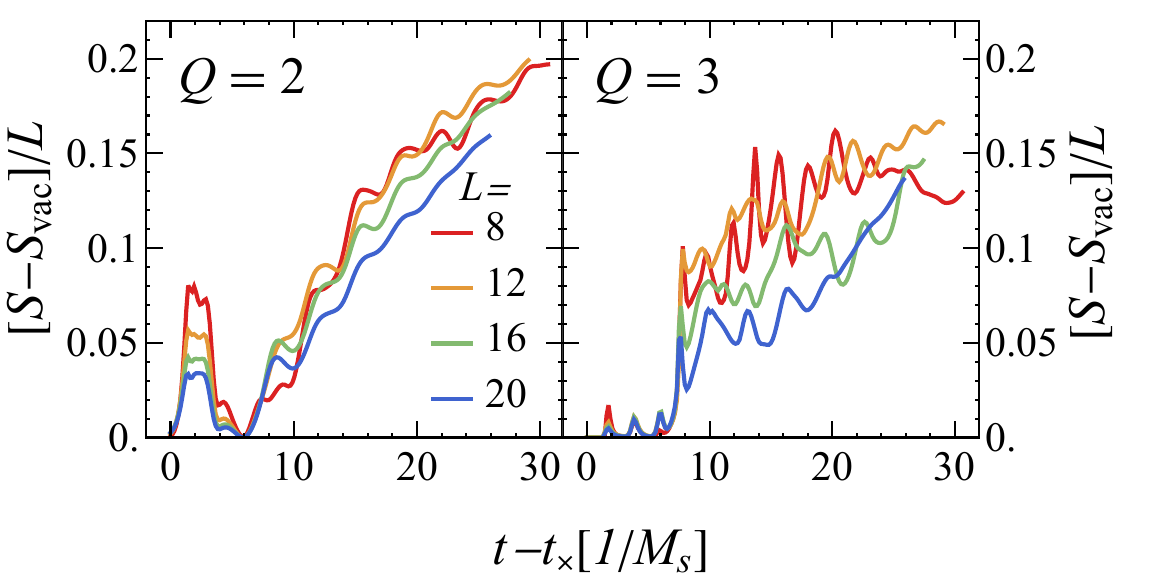}
    \caption{Emergence of volume law in the time evolution of  symmetry-resolved entanglement entropy in the charge sectors $Q=0,1,2,3$. Vacuum values of EE are subtracted, the result is rescaled by the subsystem size and time is shifted as explained in the text describing Fig. \ref{fig:S_area_vol_law}. $Q=0, 1$ sectors exhibit features similar to the full entanglement entropy, to which they provide dominant contributions (along with $Q=-1$ that is the same as $Q=1$). In $Q=2$ sector one can see that the volume law is approximately satisfied almost from the beginning. On the contrary, in $Q=3$ sector we do not find volume law in small subsystems (which can be attributed to the fact that there are only a few states with $Q=3$ in these subsystems).}
    \label{fig:Q0123_vol_law}
\end{figure}
The reduced density matrix is then written as a direct sum over the symmetry sectors:
\begin{equation}
    \rho = \bigoplus_{Q=-L/2}^{L/2} \rho_Q \,.
\end{equation}
Let us define a normalized version of $\rho_Q$, namely $\tilde \rho_Q$ such that $\Tr \,\tilde \rho_Q = 1$ with $\tilde\rho \equiv \frac{1}{\Tr\,\rho} \rho$. The entanglement spectrum undergoes the same renormalization: 
\begin{equation}
    \tilde\lambda_i^Q \equiv \frac{1}{\sum_j \lambda_j^Q} \lambda_i^Q \,. 
\end{equation}
Then, one can define the symmetry-resolved entanglement entropy in charge sector $Q$:
\begin{equation}
    S_Q = -\sum_{i} \tilde\lambda_i^Q \log \tilde \lambda_i^Q \, .
\end{equation}
Symmetry-resolved entanglement entropy has emerged in recent years as a powerful tool to study the interplay between quantum entanglement and symmetries of a theory, both in many-body systems (see e.g. \cite{Xavier:2018kqb, Turkeshi:2020yxd}) and in quantum field theories \cite{Goldstein:2017bua, Murciano:2020vgh}. 
In the context of our work, we are interested in the interplay between the $U(1)$ symmetry of the Schwinger model and thermalization. Below, we study if thermalization found in the full entanglement entropy is reproduced at the symmetry-resolved level. 

We begin by studying the emergence of volume law scaling in the evolution of symmetry-resolved entanglement entropy in the jet system. Focusing on the charge sectors $Q=0,1,2,3$ (we find that due to the symmetry of the quench the entanglement spectra in sectors $+Q$ and $-Q$ are equivalent), we present the results for the vacuum subtracted $S_Q$ rescaled by the subsystem volume in Fig. \ref{fig:Q0123_vol_law}. Volume law is clearly reached at late times in the $Q=0,1,2$ sectors. In the $Q=3$ sector the fluctuations make it challenging to reach a conclusion about volume law behavior. It may be related to the fact that the $L=8, 12$ subsystems do not possess sufficiently many states with charge $Q=3$.

\begin{figure}
    \centering
    \includegraphics[width=1\linewidth]{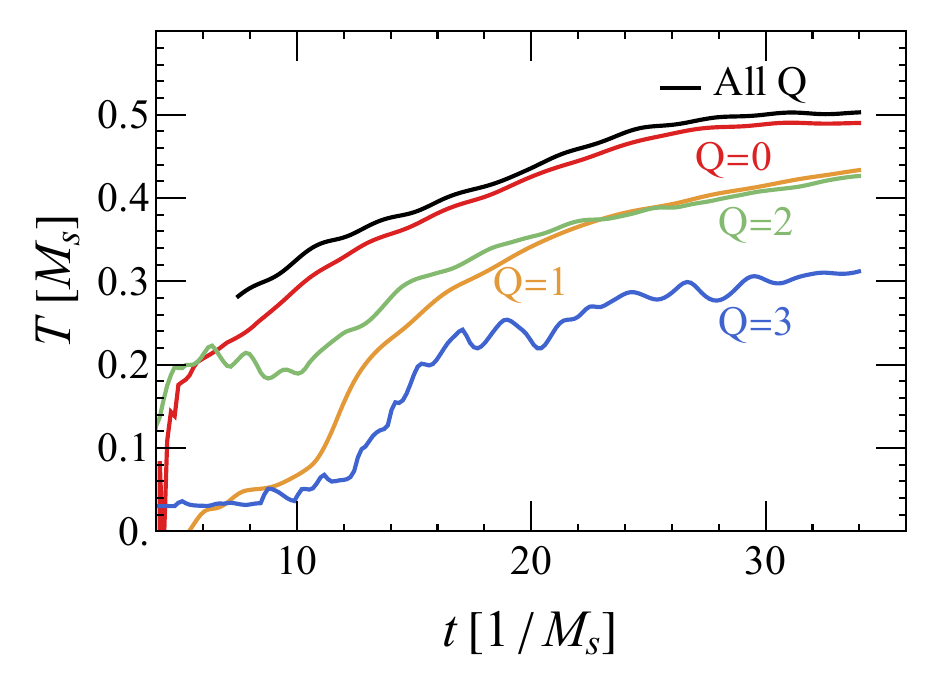}
    \caption{Effective temperature extracted from the time evolution of symmetry-resolved entanglement entropy in different charge sectors. Infinite volume extrapolated Gibbs entropy is used to extract the temperature. Black line corresponds to the full entanglement entropy and is the same as the black line in Fig. \ref{fig:thermalization_m025}. $Q=0$ symmetry-resolved temperature turns out to be very close to the full entropy temperature, while in the charged sectors the effective temperature is lower.}
    \label{fig:SymResEntTemp}
\end{figure}

 We can also study the effective temperature extracted from the symmetry-resolved entanglement by assuming an equivalence between the symmetry-resolved entanglement entropy and Gibbs entropy by charge sector, defined as
\begin{equation}
    S_{\rm Gibbs}^Q = -\sum p_n^Q \log p_n^Q,
\end{equation}
where $p_n^Q = e^{-E_n^Q/T}/Z_Q$ are the (renormalized) Boltzmann weights in the charge $Q$ sector; $Z_Q \equiv \sum_n e^{-E_n^Q/T}$. As before with the full Gibbs entropy, we access it from exact diagonalization for a variety of (small) system sizes $N$ and perform a linear extrapolation in $1/N$. Using the inverse of $S^Q_{\rm Gibbs}(T)_{N\rightarrow\infty}$ as $T(S^Q)$, we extract the effective temperature from each symmetry sector from a sample of subsystems with $L\in[8,12]$ as was done in Sec.~\ref{sec:entropy} for the full entropy . The results are presented in Fig. \ref{fig:SymResEntTemp}.

We note that in the charge-0 sector the effective temperature follows closely the one obtained from the full entanglement entropy.  While in $Q=1,2,3$ sectors the temperature also increases with time, it is lower than the temperature in $Q=0$ sector. It would be very interesting to find a physical reason for this, and we leave it for future work.

\section{Overlaps of density matrices at $L=2$} \label{app:overlaps_L_2}

\begin{figure}
    \centering
    \includegraphics[width=\linewidth]{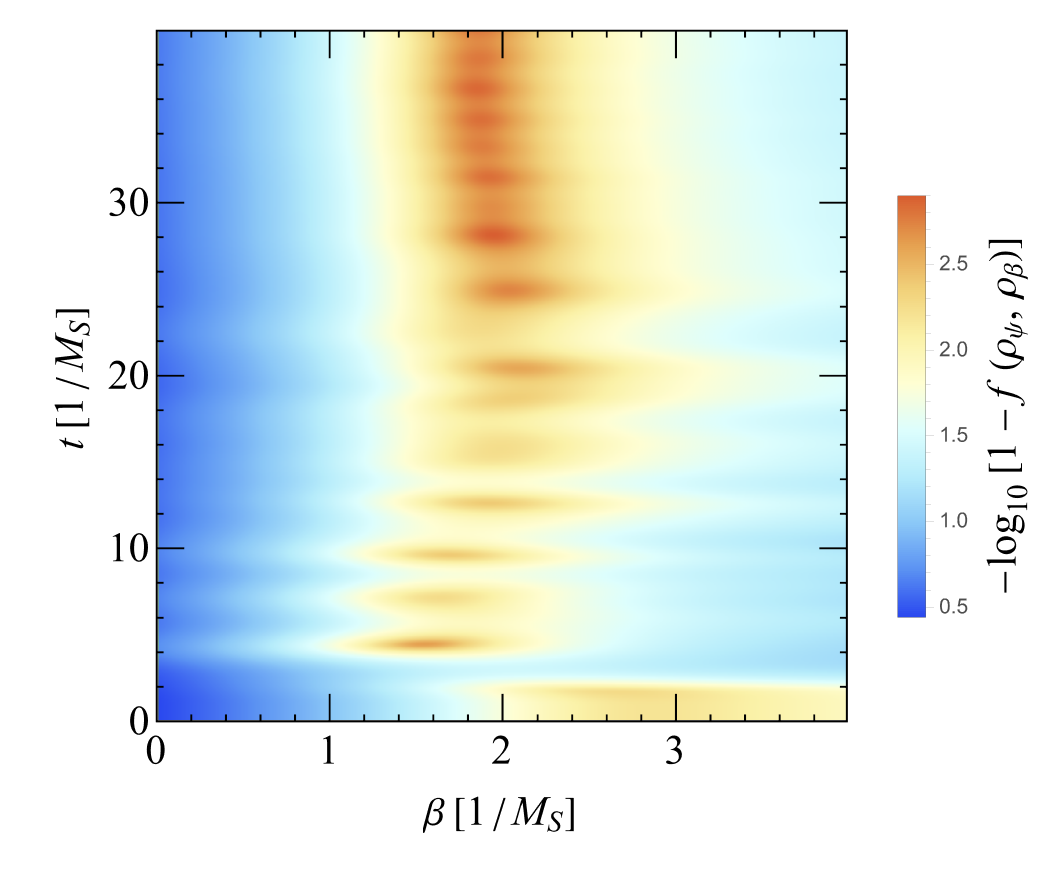}
    \caption{Overlap of a reduced density matrix obtained from the jet system state $|\Psi(t)\rangle$ for a subsystem with $L=2$ (namely, the subsystem consisting of sites number 49 and 50) with a thermal state $\rho_\beta$ as function of $t$ and $\beta$, shown as a logarithm of deviation from 1 for visual clarity.}
    \label{fig:overlap_density}
\end{figure}

In a small subsystem with $L=2$ the density matrix can be expanded in a complete basis of local operators with an intuitive physical meaning. The corresponding Hilbert space is 4-dimensional and thus the density matrix is $4\times 4$. However, the presence of a conserved electric charge yields a significant simplification: the density matrix becomes block-diagonal with a $2\times 2$ block corresponding to the charge-0 sector and two single diagonal entries, corresponding to charge $\pm1$. Thus the density matrix for a subsystem consisting of sites $n$ and $n+1$ can be expanded in a complete basis of 6 local operators, namely:
\begin{align}
   \hat{\cal O}^{(2)}=\{&\frac{1}{2}, \frac{\widehat{\bar\psi\psi}_n + \widehat{\bar\psi\psi}_{n+1}}{\sqrt{2}}, \frac{Q_n+Q_{n+1}}{\sqrt{2}}, \nonumber \\ &2\widehat{\bar\psi\psi}_n \widehat{\bar\psi\psi}_{n+1}, \sqrt{2}K_n, \sqrt{2}J_n\} \,,
\end{align}
where to complete the basis we introduced the local electric current operator $J_n\equiv -\frac{1}{2}(\chi_n^\dagger \chi_{n+1} + \chi_{n+1}^\dagger \chi_n)$. The normalization of the operators is chosen so that they form an orthonormal basis:
\begin{equation}
    \Tr(\hat{\cal O}^{(2)}_i \hat{\cal O}^{(2)}_j) = \delta_{ij} \,.
\end{equation}
Then the coefficients of the expansion of the subsystem density matrix are given by the expectation values of the  operators in the corresponding state:
\begin{equation}
    \rho = \sum_{i=1}^6 \langle {\cal O}^{(2)}_i\rangle_\rho \hat{\cal O}^{(2)}_i \,.
\end{equation}
The overlap between the reduced density matrix in a jet system $\rho_\psi$ and a thermal density matrix in a system with $N=2$ sites $\rho_\beta$ can now be computed as
\begin{equation}
    f(\rho_\psi, \rho_\beta) = \frac{\sum_{i=1}^6\langle {\cal O}^{(2)}_i\rangle_\psi \langle {\cal O}^{(2)}_i\rangle_\beta}{\sqrt{\sum_{j=1}^6(\langle {\cal O}^{(2)}_j\rangle_\psi)^2} \sqrt{\sum_{k=1}^6(\langle {\cal O}^{(2)}_k\rangle_\beta)^2}} \,. \label{eq:overlap_from_basis}
\end{equation}
Note that due to charge-neutral and time-independent nature of a thermal state $\langle Q_1 + Q_2\rangle_\beta = \langle J_1\rangle_\beta = 0$, thus in the above equation indices $i$ and $k$ do not assume values 3 and 6.

\begin{figure}
    \centering
    \includegraphics[width=1\linewidth]{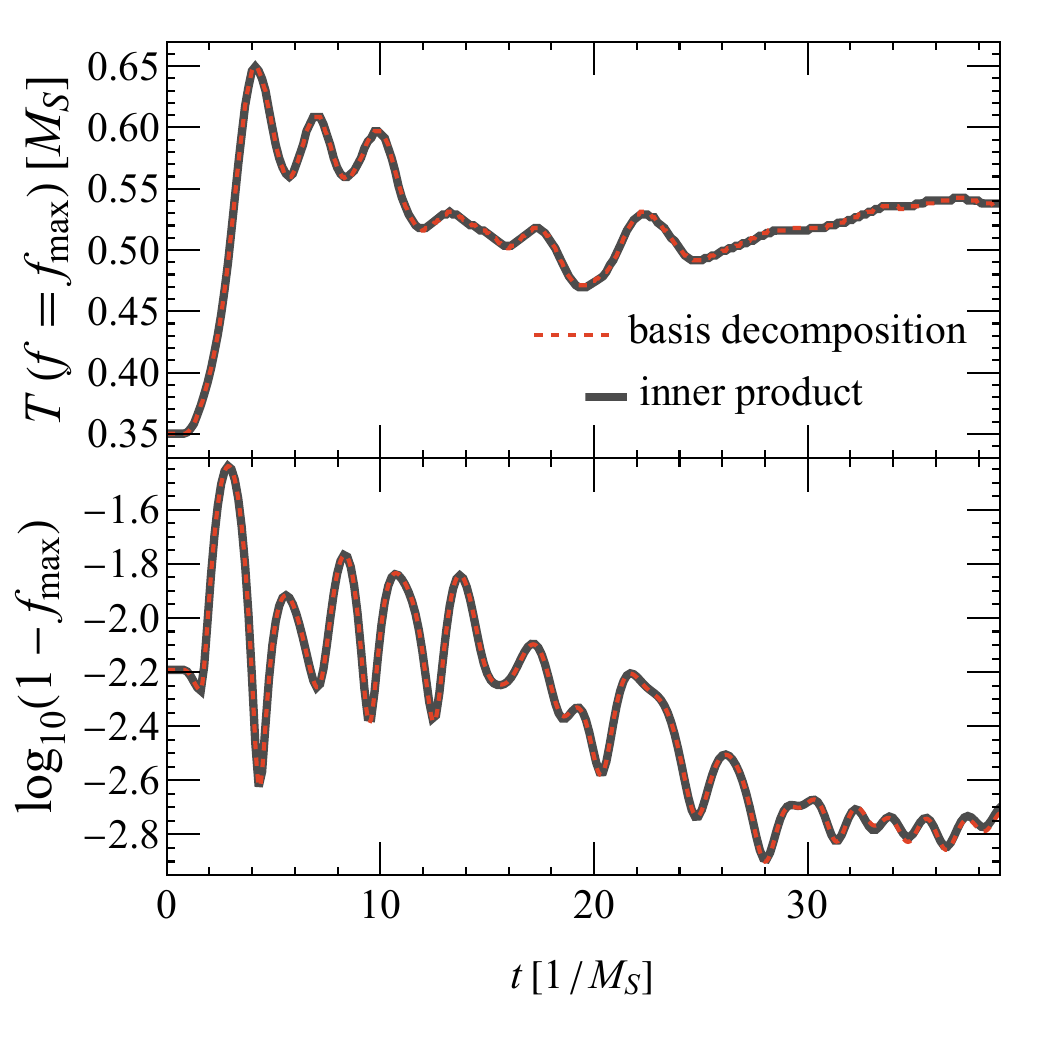}
    \caption{Top panel: time dependence of temperature of the thermal state with maximal overlap with the  reduced density matrix for a jet system describing subsystem of sites number 49 and 50. Bottom panel: the log of difference between the maximal overlap and 1. In both panels, the dashed red lines refer to the overlaps found from comparing local observables, as in Eq. (\ref{eq:overlap_from_basis}). The solid black lines refer to the overlaps found from Eq. (\ref{eq:overlap_definition}) by taking inner products of MPOs, see the discussion in Section \ref{sec:overlap} of the main text. The two methods produce identical results, providing a cross-check of the validity of both.}
    \label{fig:Overlaps_2_site_comparison}
\end{figure}

In Fig.~\ref{fig:overlap_density} we display how the overlap of an evolving jet subsystem (that consists of sites number 49 and 50) with a thermal state depends on time and on temperature of the thermal state. At late times the maximal values of overlap reach values very close to 1. The maximal overlap and the temperature of the corresponding thermal state are shown in Fig.~\ref{fig:Overlaps_2_site_comparison} where they are also compared to the values found for the same subsystem with the method used in the main text, namely by contracting the MPOs directly instead of using local observables. The two methods are in perfect agreement.

Note that while a complete understanding of the density matrix in terms of local observables with a clear physical meaning is feasible for $L=2$ subsystem, it becomes impractical for larger subsystems. In particular, for $L=4$ the density matrix is $16\times 16$ and consists of 5 blocks in different charge sectors, namely $6\times 6$ in charge 0, $4\times 4$ in charge $\pm 1$ and single entries in charge $\pm 2$. Overall, 70 local operators are necessary to decompose a general $L=4$ density matrix. While it may be possible to assign physical meaning to all these operators, it is a tedious task that we leave for future work. Computing the overlap via a product of MPOs does not require understanding of the basis decomposition of density matrices, and thus we employ this method for systems with $L\geq4$ in the main text.

\section{$\cal CP$-symmetric Hamiltonian in general charge sector} \label{app:L_charge_Q}

The Gauss's law solution for the electric field given in the second equation of Eq.(\ref{eq:Gauss_law}) is defined with the total charge $Q=0$ in mind. Evidently at a nonzero net charge in this definition there is an asymmetry between the zero electric flux entering the system on the left and flux $Q$ exiting it on the right. This breaks the $\cal CP$ symmetry of the Hamiltonian. In order to restore it, we redefine the electric field as
\begin{equation} \label{eq:L_with_Q_tot}
    L_n = \sum_{i=1}^n Q_i - \frac{1}{2}\sum_{i=1}^N Q_i = \frac{1}{2}\left[\sum_{i=1}^n Q_i -\sum_{i=n+1}^N Q_i \right] \,.
\end{equation}
Note that it coincides with the definition given in Eq.(\ref{eq:Gauss_law}) if $\sum_{i=1}^N Q_i = 0$. 

One can further simplify the electric field term in the Hamiltonian $H_E = \frac{ag^2}{2}\sum_{i=1}^{N-1}L_n^2$ that involves triple sums when $L_n$ is plugged in from Eq.(\ref{eq:L_with_Q_tot}). In the spin formulation, described in Appendix \ref{app:spin}, one of the summations can be carried out analytically, leading to the following expression convenient for implementation:
\begin{align}
    H_E = \frac{ag^2}{16}\bigg(\frac{N(N+1)}{2}+ (N-1)\sum_{i=1}^N (-1)^i Z_i  + \nonumber\\
    \sum_{i=2}^N\sum_{j=1}^{i-1} (N-1-2i+2j)\big[Z_iZ_j + (-1)^iZ_j 
    + (-1)^jZ_i\big]\bigg).
\end{align}

\vspace{1cm}

\FloatBarrier

\bibliographystyle{utphys}
\bibliography{main}

\clearpage

\end{document}